\documentclass[sigconf,nonacm]{acmart}

\usepackage{booktabs}
\usepackage{braket}
\usepackage{graphicx}
\usepackage{algorithm}
\usepackage[noend]{algpseudocode}
\usepackage{tikz}
\usepackage{pifont}
\usetikzlibrary{arrows.meta,positioning,fit,calc}

\title{
  Do Not Let CNOTs Overwhelm the Decoder:\\
  Scheduling Transversal Gates for Fast FTQC
}
\renewcommand{\shorttitle}{
  Do Not Let CNOTs Overwhelm the Decoder: Scheduling Transversal Gates for Fast FTQC
}

\setcopyright{none}
\usepackage[most]{tcolorbox}
\newtcolorbox{keyobservation}{
    enhanced,
    breakable,
    colback=white,
    frame hidden,
    boxsep=0pt,
    left=5pt,
    right=0pt,
    top=0pt,
    bottom=0pt,
    before skip=2pt,
    after skip=2pt,
    overlay={
        \draw[black!45, line width=1pt]
        ([yshift=-2pt]frame.north west)
        --
        ([yshift=2pt]frame.south west);
    },
}

\author{Shota Ikari}
\affiliation{%
  \institution{The University of Tokyo / RIKEN}
  \city{Tokyo}
  \country{Japan}
}
\email{shota-ikari@g.ecc.u-tokyo.ac.jp}

\author{Yuga Hirai}
\affiliation{%
  \institution{Keio University / RIKEN}
  \city{Kanagawa}
  \country{Japan}
}
\email{yuga917@keio.jp}

\author{Yasunari Suzuki}
\affiliation{%
  \institution{RIKEN}
  \city{Saitama}
  \country{Japan}
}
\email{yasunari.suzuki@riken.jp}

\author{Hiroshi Nakamura}
\affiliation{%
  \institution{The University of Tokyo}
  \city{Tokyo}
  \country{Japan}
}
\email{nakamura@hal.ipc.i.u-tokyo.ac.jp}

\author{Yosuke Ueno}
\affiliation{%
  \institution{RIKEN / The University of Tokyo}
  \city{Saitama}
  \country{Japan}
} 
\email{yosuke.ueno@riken.jp}

\begin{abstract}
Transversal CNOT~(TCNOT) gates can accelerate fault-tolerant quantum computation~(FTQC) in the surface code by reducing the number of syndrome extraction rounds required between logical operations from $O(d)$ to $O(1)$.
This is particularly attractive for quantum platforms with long-range connectivity, such as neutral-atom arrays.
However, dense TCNOT schedules substantially increase the classical decoding workload.
Error propagation through TCNOTs correlates detection events across multiple surface-code patches, enlarging the spatiotemporal region that must be decoded jointly.
Consequently, denser TCNOT schedules increase decoding latency and memory requirements and potentially exceed available decoder capacity. 
Moreover, because the detector error model (DEM) of each decoding window depends on the TCNOT schedule, exhaustively precomputing all possible window-level DEMs is infeasible, requiring just-in-time (JIT) DEM compilation. 
Thus, the practical benefit of TCNOT gates is limited not only by quantum hardware performance but also by classical decoding and DEM-compilation capacity.

We introduce \emph{PACE}, a decoder-aware TCNOT scheduling framework for surface-code systems supporting TCNOT gates. 
PACE first mitigates the decoder-side costs of aggressive TCNOT scheduling through three complementary techniques.
\emph{Hybrid Window Decoding} assigns different decoders for each decoding window according to its DEM structure.
\emph{DEM Stitch} generates schedule-specific window-level DEMs just in time by assembling reusable precompiled fragments. 
\emph{Sub-window Parallel Decoding} decomposes large decoding windows into smaller sub-windows and uses a graph-coloring formulation to expose and schedule parallel decoding opportunities. 
Building on these techniques, PACE then performs decoder-aware scheduling to maximize TCNOT concurrency within the available decoder resources.
Our evaluation characterizes the trade-off between quantum acceleration and classical decoding cost, revealing the limitations of current decoding systems for TCNOT-based FTQC and highlighting directions for future decoder development.
\end{abstract}

\begin{document}
\maketitle

\section{Introduction}
Fault-tolerant quantum computing (FTQC) enables quantum algorithms to run at low logical error rates by encoding logical information in quantum error-correcting (QEC) codes~\cite{Nielsen_Chuang_2010, gottesman2010introduction}.
Among the many proposed QEC codes, the surface code is one of the leading candidates because it requires only nearest-neighbor stabilizer measurements and has a high error threshold~\cite{dennis2002topological,fowler2012surface, acharya2025belowthreshold}.

During quantum computation, imperfect qubits and gates continuously introduce physical errors. 
Syndrome extraction (SE) repeatedly measures stabilizers to detect these errors, producing syndrome data that must be decoded in real time.
For online decoding, the stream of syndrome measurements is partitioned into overlapping decoding windows~\cite{dennis2002topological,skoric2023parallel}, each decoded using its corresponding detector error model (DEM), a graph-based representation of the decoding problem~\cite{gidney2021stim}.
To sustain real-time operation, decoding must keep pace with SE to prevent the backlog problem~\cite{terhal2015quantum,battistel2023real}.

Recent studies have proposed transversal logical operations to reduce the quantum execution time of surface-code computation~\cite{chen2026transversal, chen2026hierarchical, zhou2025low, zhou2025resourceanalysis}.
Whereas lattice-surgery-based logical operations typically require $O(d)$ rounds of SE, transversal CNOT (TCNOT) gates require only $O(1)$ rounds, where $d$ denotes the code distance~\cite{bombin2023logical, fowler2018low}.
This is advantageous for quantum platforms with long-range connectivity, such as neutral-atom arrays~\cite{bluvstein2025fault}, enabling TCNOTs to substantially reduce quantum execution time.

\begin{figure*}[t]
    \centering
    \includegraphics[width=0.9\linewidth]{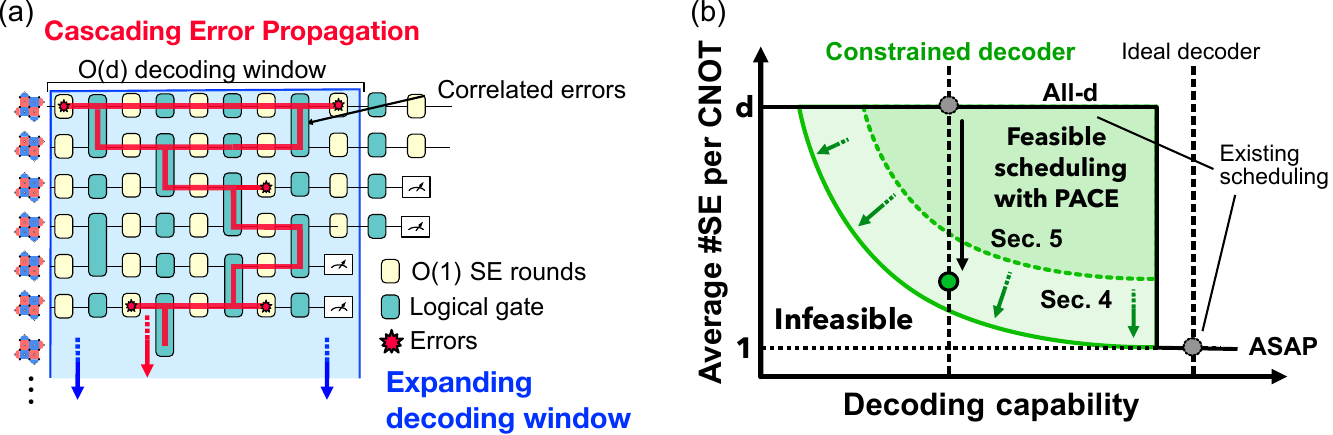}
        \caption{
        Motivation of PACE.
        (a) Dense TCNOT scheduling enlarges the decoding window, rapidly increasing decoder workload.
        (b) PACE schedules TCNOTs as densely as decoder capacity allows, balancing quantum acceleration with decoding capability.
        }
    \label{fig:deft_system_overview}
\end{figure*}

However, existing studies have mainly focused on the quantum-side acceleration provided by TCNOTs, while their impact on the classical decoding system has received much less attention~\cite{sunami2025transversalgame}.
TCNOTs introduce three fundamental challenges for online decoding.
First, TCNOTs create correlated errors across multiple surface-code patches.
As a result, a single physical error can flip more than two detectors, appearing as a hyperedge in the corresponding DEM.
Such DEMs are no longer purely graphlike~\cite{higgott2025sparse} and non-graphlike DEMs require more complex decoding algorithms~\cite{cain2025fast,cain2024correlated,wan2024iterative}, which increase decoder latency and runtime memory usage.
Second, TCNOT scheduling exponentially increases the diversity of possible window-level DEMs.
Exhaustively preparing DEMs for every possible schedule is therefore impractical, requiring DEMs to be generated online before decoding can proceed.
This online DEM preparation~\cite{ziad2026greenpeas} is frequently overlooked, yet it can itself become a significant bottleneck for real-time decoding.
Third, aggressive TCNOT scheduling can rapidly enlarge the decoding window~\cite{sahay2025error,zhou2025resourceanalysis}, increasing both decoder latency and working-memory requirements.
Fully exploiting the quantum-side acceleration of TCNOTs would require decoding arbitrarily large correlated volumes within fixed latency and memory constraints, which is infeasible because aggressive TCNOT scheduling rapidly increases both the decoding volume and its associated costs.

Existing approaches make different tradeoffs between TCNOT concurrency and classical decoding complexity.
Sahay \textit{et al.}~\cite{sahay2025error} keep each correlated decoding problem manageable by restricting a decoding block to a limited number of TCNOTs and separating successive blocks with sufficient SE rounds.
This approach limits the size and diversity of DEMs, but reduces TCNOT concurrency and thus gives up part of the potential quantum-side acceleration.
Cain \textit{et al.}~\cite{cain2025fast,cain2024correlated} develop correlated-decoding methods that support more aggressive TCNOT execution by reducing or reformulating the associated decoding work.
These methods focus on decoding a given transversal circuit, rather than selecting its TCNOT schedule under system-level decoder constraints.
Consequently, existing work does not jointly consider online DEM preparation, per-window memory and decoding-latency limits, and their dependence on the TCNOT schedule.
This gap motivates selecting the TCNOT schedule according to both quantum execution time and the capacity of the classical decoding system.

This paper proposes \emph{PACE}, a scheduling framework for surface-code systems with transversal CNOTs.
PACE first mitigates the decoder-side costs of TCNOT by combining three techniques: \emph{hybrid window decoding} to reduce decoding latency and working memory, \emph{just-in-time (JIT) DEM compilation} to construct window-level DEMs at runtime and \emph{sub-window parallel decoding} to manage volume of each decoding window.
These techniques directly address the three challenges.
PACE then leverages these techniques to perform decoder-aware scheduling that maximizes quantum-side acceleration while maintaining decoder feasibility~(Fig.~\ref{fig:deft_system_overview}(b)).
Our contributions are as follows:
\begin{itemize}
\item We identify three decoder-side bottlenecks introduced by aggressive TCNOT scheduling: (i)~correlated-unmatchable decoding cost, (ii)~the growing diversity of DEM patterns, and (iii)~decoding volume growth~(Sec.~\ref{sec:motivation}).
\item We present \emph{PACE}, a decoder-aware framework for FTQC with TCNOT that mitigates these bottlenecks through three complementary techniques: hybrid window decoding, JIT DEM compilation, and sub-window parallel decoding.
These techniques motivates decoder-aware scheduling that balances quantum acceleration with classical decoding capability~(Sec.\ref{sec:mitigation}, \ref{sec:scheduling}).
\item We evaluate PACE using decoder evaluations, DEM construction experiments, and benchmark quantum circuits, demonstrating that it reduces quantum execution time while satisfying the decoder constraints~(Sec.\ref{sec:setup}, \ref{sec:results}).
\end{itemize}

\section{Background}
\label{sec:background}

\subsection{Surface Code}
\begin{figure*}[t]
    \centering
    \includegraphics[width=0.8\linewidth]{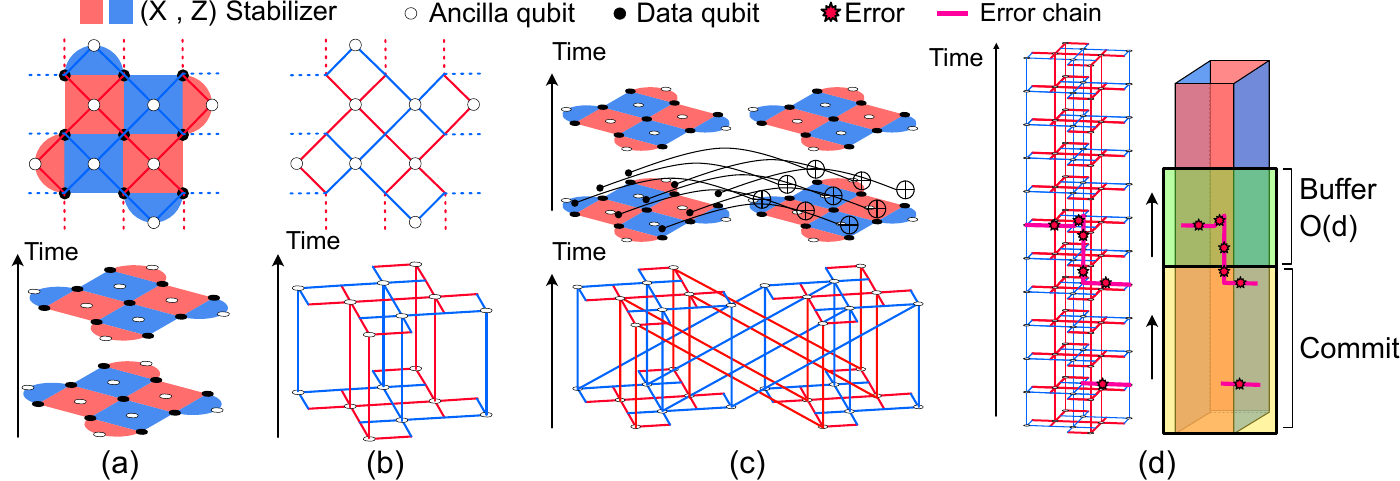}
    \caption{(a) A distance-three rotated surface code.
            (b) A DEM for a memory experiment.
            (c) A TCNOT can flip more than two detectors, which forms hyperedges in the corresponding DEM.
            (d) The diagram of window decoding.}
    \label{fig:DEM}
    
\end{figure*}

The surface code is one of the most widely studied QEC codes because it combines a high error threshold with stabilizer measurements that require only nearest-neighbor interactions on a two-dimensional lattice~\cite{bravyi1998quantum,dennis2002topological,fowler2012surface}.
As a CSS code, it employs separate $X$- and $Z$-type stabilizers to detect phase-flip and bit-flip errors, respectively~(Fig.~\ref{fig:DEM}(a)).
SE repeatedly measures these stabilizers without directly disturbing the encoded logical state, producing a record of syndrome measurements.
We denote the duration of one SE round by $\tau_{\mathrm{SE}}$.
Detectors are computed as parities of selected syndrome measurements in this record.
A detection event occurs when a detector's outcome differs from its expected value, indicating a possible physical fault.
During execution, classical decoders continuously process the resulting stream of detection events to infer likely errors on the quantum device, and they must keep pace with its generation.
In conventional surface-code error correction, the resulting DEM is typically graphlike, allowing efficient decoding using matching-based decoders~\cite{higgott2025sparse, griffiths2024union}.

\subsection{Transversal CNOT (TCNOT) Gate}
A TCNOT gate is a logical CNOT gate implemented by applying physical CNOT gates between corresponding data qubits in two surface-code patches~\cite{gottesman2010introduction}.
This operation requires long-range connectivity between corresponding data qubits in the two surface code patches, making it well suited to reconfigurable architectures such as neutral-atom quantum computers~\cite{bluvstein2023logical,bluvstein2025fault}.
Recent works have shown that, under suitable decoding assumptions, TCNOT gates can be sequentially scheduled with only $O(1)$ SE rounds between other logical operations~\cite{zhou2025low,zhou2025resourceanalysis}.
This is substantially smaller than the $O(d)$ rounds required by lattice-surgery~\cite{horsman2012lattice, fowler2018low}; $d$ denotes the code distance.

\subsection{Detector Error Models (DEMs)}
\label{sec:dem}
The decoder requires prior knowledge of the noise characteristics of the quantum device and the executed circuit to estimate errors correctly.
This information is represented by a DEM~\cite{gidney2021stim}, a graph-based model that characterizes the relationship between physical errors, detectors, and logical observables.
A detector is a parity check over syndrome measurements whose value is deterministic in the absence of errors and is represented as a node in the DEM.
Each physical error is represented as an edge connecting the detectors whose outcomes are flipped by that error, while logical observables are associated with the corresponding edges.

Following~\cite{beverland2025fail}, we represent a DEM as $D=(p,H,A)$, where $p\in[0,1]^N$ is the fault-probability vector, $H\in\mathbb{F}_2^{M\times N}$ is the detector check matrix, and $A\in\mathbb{F}_2^{K\times N}$ is the logical-action matrix.
Each fault mechanism corresponds to one column of $H$ and $A$, together with its probability in $p$.
If a fault vector $e\in\mathbb{F}_2^N$ occurs, it produces the detection events $\sigma=He$.
A decoder maps the observed detection events $\sigma$ to an estimate $\hat e$ satisfying $H\hat e=\sigma$.
Decoding succeeds if $A\hat e=Ae$, where all operations are over $\mathbb{F}_2$.

Figure~\ref{fig:DEM}(b) illustrates this representation for a surface-code memory experiment~\cite{gidney2022stability} and its DEM.
When every fault mechanism flips at most two detectors, the resulting DEM is graphlike and can be decoded efficiently using matching-based decoders~\cite{higgott2025sparse,delfosse2021almost,griffiths2024union}, which we refer to as \emph{matchable} decoders.
In contrast, TCNOT gates propagate Pauli errors between the control and target patches, allowing a single fault to flip more than two detectors across multiple surface-code patches.
As illustrated in Fig.~\ref{fig:DEM}(c), the resulting hyperedges break the graphlike structure of the DEM and require decoders capable of handling correlated faults, which we refer to as \emph{unmatchable} decoders~\cite{hillmann2025localized, muller2025improved, beni2025tesseract, roffe2020decoding}.

\begin{figure}[t]
    \centering
    \includegraphics[width=\linewidth]{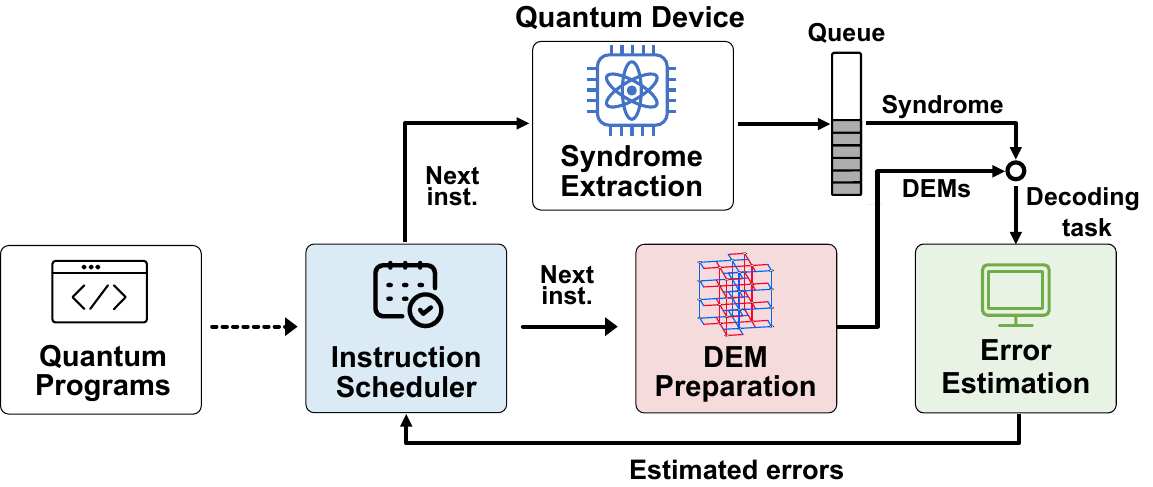}
    \caption{Execution pipeline of the FTQC system considered in this study.
        The highlighted components indicate the scope of this work.}
    \label{fig:pipeline}
\end{figure}

\subsection{Window Decoding}
\label{sec:window_decoding}
Window decoding is an online decoding scheme in which the continuous stream of detection events is decomposed into overlapping windows, each decoded using its corresponding DEM while the quantum device continues producing later syndrome measurements~\cite{dennis2002topological,skoric2023parallel, chen2026triage,viszlai2024predictive, toshio2025decoder}.
To prevent the degradation of decoding performance caused by dividing the window, a commit region and a buffer region are often introduced, as shown in Fig.~\ref{fig:DEM}~(d).
As decoding proceeds, windows slide forward in time while overlapping through their buffer regions.
The buffer region should have $O(d)$ rounds to preserve the code distance $d$ across the commit region, ensuring that errors outside the window do not affect the detectors in the commit region.
Throughout this paper, we set the buffer region to $W_B = d$ rounds, and denote the window size by $W$.
The commit-region length, and hence the window stride, is therefore $W_{\mathrm{C}}=W-d$ rounds.

\subsection{FTQC Computational Flow}
Figure~\ref{fig:pipeline} illustrates the execution pipeline of an FTQC system.
The instruction scheduler issues logical instructions according to circuit dependencies and decoded feedback.
While the quantum device continuously performs SE, the classical pipeline prepares window-level DEMs and decodes the resulting detection events.
The estimated errors are fed back to the scheduler to support classically controlled operations.
Because the quantum and classical pipelines operate concurrently, FTQC throughput is limited by the slower pipeline.
DEM preparation and decoding must therefore sustain the rate at which SE generates new decoding tasks; otherwise, the tasks accumulate, delaying decoder feedback and eventually preventing real-time FTQC execution~\cite{terhal2015quantum}.

\subsection{Instruction Set for FTQC with Transversal Gates}
\label{sec:instruction_set}
\begin{figure}[t]
    \centering
    \includegraphics[width=\linewidth]{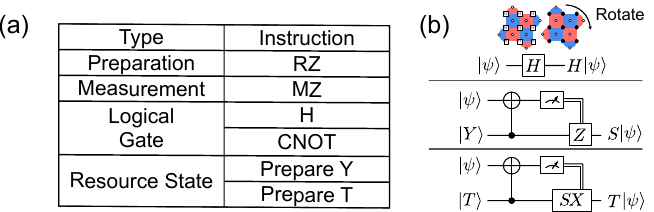}
    \caption{(a) Instruction set assumed throughout this work.
    (b) Implementation of logical gates.
    Logical $H$ is realized by a direct patch rotation following the transversal $H$ gate;
    Logical $S$ and $T$ are implemented by magic-state teleportation using $|Y\rangle$ and $|T\rangle$, respectively. }
    \label{fig:logical_operations}
    
\end{figure}

We assume a surface-code architecture supporting TCNOT gates for Clifford+$T$ computation, following~\cite{sunami2025transversalgame}.
Our focus is on the decoder-side cost of TCNOT-based logical operations; quantum-circuit optimization is beyond the scope of this work~\cite{wang2026transpiler}.
Figure~\ref{fig:logical_operations} summarizes the instruction set and logical-gate implementations in this study.
Under the assumed instruction set, logical CNOT gates are implemented by TCNOTs between aligned logical patches, $H$ gates by direct patch rotation~\cite{cicali2025fast,sunami2025transversalgame}, and $S$ / $T$ gates by magic-state teleportation~\cite{litinski2019game}.
The $S$ and $T$ gadgets consist of CNOT operations, measurements, and the corresponding Pauli-frame updates~\cite{chamberland2018pauli,riesebos2017pauli}.
Note that our proposal is not limited to this specific instruction set and applies to any FTQC architecture supporting transversal CNOT operations.

\section{Motivation and PACE Overview}
\label{sec:motivation}
Although dense TCNOT scheduling substantially accelerates quantum execution, it also increases the computational burden on the classical decoder.
We first observe the prevalence of TCNOTs in the instruction set.
Then, the following subsections describe the three major decoder-side challenges introduced by dense TCNOT placement.
Finally, we overview how our proposed framework PACE addresses them.

\subsection{Prevalence of TCNOTs in FTQC}
\begin{figure}
    \centering
    \includegraphics[width=\linewidth]{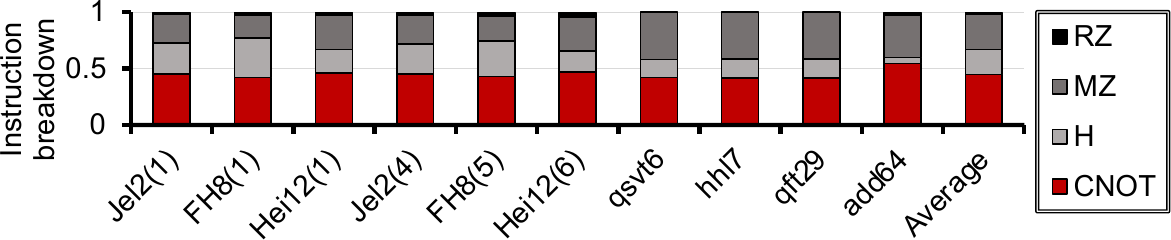}
    \caption{Instruction breakdown of the benchmark circuits in Table~\ref{tab:benchmark_characterization}, using the instruction set in Fig.~\ref{fig:logical_operations}. The resource-state preparations are omitted because they are not primitive operations and do not lie on the execution critical path.} 
    \label{fig:instruction_breakdown}
    
\end{figure}

Using the instruction set in Fig.~\ref{fig:logical_operations}\,(a), $S$ and $T$ gates in a given FTQC program are performed with magic-state teleportation, which introduces additional TCNOT operations, as shown in Fig.~\ref{fig:logical_operations}\,(b).
Consequently, TCNOTs constitute a substantial fraction of the FTQC program execution.
To quantify this observation, Fig.~\ref{fig:instruction_breakdown} presents the instruction breakdown of the benchmark circuits summarized in Table~\ref{tab:benchmark_characterization}.
Across these benchmarks, TCNOTs account for 40--55\% of all primitive quantum instructions.
This high prevalence of TCNOT gates motivates the decoder-side challenges due to TCNOTs discussed in the following subsections.

\subsection{Correlated Errors with TCNOTs}
\label{sec:hyperedge}
\begin{figure}[t]
    \centering
    \includegraphics[width=\linewidth]{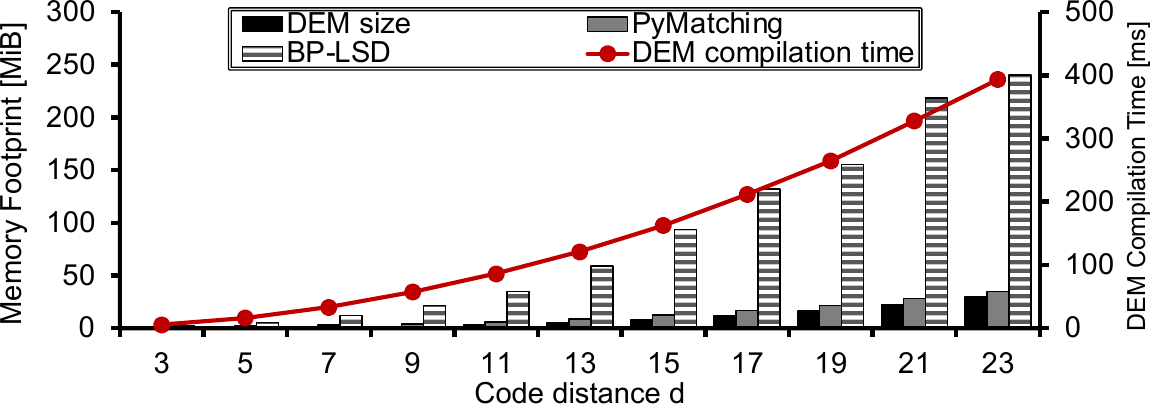}
    \caption{The memory footprint of DEM construction and DEM preparation time measured with Stim~\cite{gidney2021stim} for surface-code memory experiments (rounds=$2d$).
    PyMatching~\cite{higgott2025sparse} and BP-LSD~\cite{hillmann2025localized} are used as examples of matchable and unmatchable decoders, respectively.
    See Sec.~\ref{sec:setup} for the experimental setup.
    }
    \label{fig:dem_memory_problem2}
    
\end{figure}

While the window-level DEM of surface-code memory experiments~\cite{gidney2022stability} is graphlike, a TCNOT makes the DEM non-graphlike by introducing correlated errors across logical patches.
Thus, in order to decode DEMs with TCNOTs, we must use unmatchable decoders that can handle hyperedges in the DEM~\cite{hillmann2025localized,beni2025tesseract,sahay2025error}.
In general, these decoders are slower and more memory-expensive than matchable decoders due to their algorithmic complexity and the data structures needed to represent hyperedges.
Figure~\ref{fig:dem_memory_problem2} compares working memory of PyMatching~\cite{higgott2025sparse} and BP-LSD~\cite{hillmann2025localized}.
Consequently, TCNOTs drastically increase decoding loads.

\subsection{The Variation of Window-Level DEMs}
\label{sec:dem_patterns}
As shown in Fig.~\ref{fig:pipeline}, timely DEM preparation is critical to sustaining the FTQC pipeline.
Constructing DEMs requires propagating circuit-level faults through the quantum circuit to determine the detectors and logical observables affected by each fault~\cite{gidney2021stim}, incurring a non-negligible latency.
As shown in Fig.~\ref{fig:dem_memory_problem2}, DEM construction using Stim~\cite{gidney2021stim} takes tens to hundreds of milliseconds, much longer than the typical SE period of a neutral-atom platform (e.g., 1~ms).
A naive approach to ensuring that the corresponding DEM is ready when each decoding task begins is to precompile all possible window-level DEMs for a given program.
However, aggressive TCNOT scheduling exponentially increases the number of possible window-level DEMs; a window spanning $O(d)$ SE rounds contains $O(d)$ candidate TCNOT slots at $O(1)$-round intervals, and the presence or absence of a TCNOT in each slot causes the number of possible window-level DEMs to grow exponentially with code distance $d$.
Moreover, as shown in Fig.~\ref{fig:dem_memory_problem2}, each DEM has a non-negligible memory footprint. 
Consequently, exhaustive DEM precompilation is impractical, making low-latency JIT DEM compilation essential.

\subsection{Increasing Volume of Decoding Window}
\label{sec:decoding_volume}
\begin{figure}[t]
    \centering
    \includegraphics[width=\linewidth]{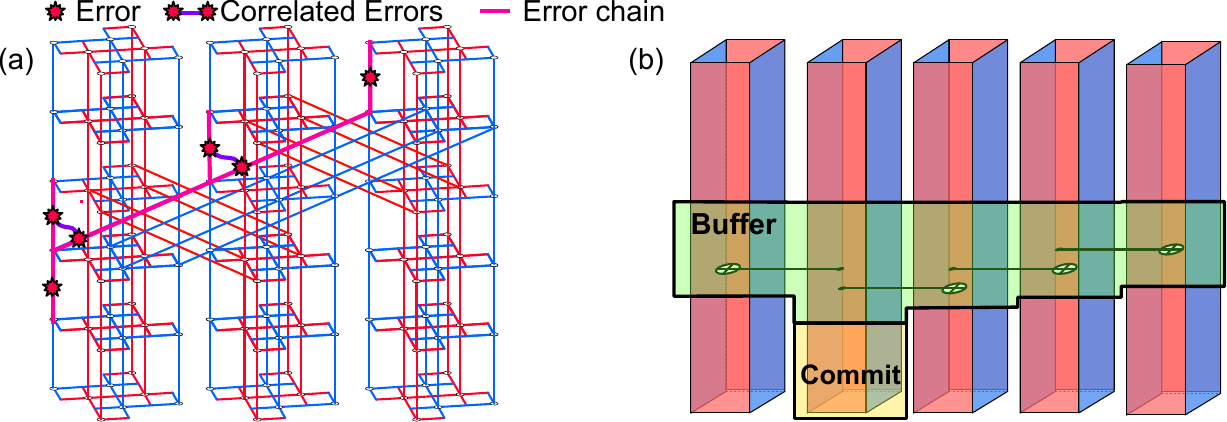}
    \caption{
    (a) TCNOT creates correlated, non-graphlike DEMs across surface-code patches.
    (b) Dense TCNOT placement increases the decoding volume within a window rapidly.}
    \label{fig:motivation_bottleneck}
    
\end{figure}

As explained in Sec.~\ref{sec:dem}, TCNOTs can propagate errors across multiple surface-code patches.
As shown in Fig.~\ref{fig:motivation_bottleneck}, aggressive TCNOT scheduling can cause error correlations induced by multiple TCNOTs within a single decoding window, substantially increasing the volume that must be decoded jointly~\cite{cain2024correlated,sahay2025error}.
Because practical decoders have finite memory capacity and processing throughput, the resulting decoding workload may exceed their real-time processing capability.
A straightforward mitigation is to separate consecutive TCNOTs by $d$ SE rounds, thereby bounding the decoding volume of each window~\cite{sahay2025error}.
However, this restriction reduces TCNOT concurrency and sacrifices much of the quantum-side speedup by aggressive scheduling.

\subsection{Strategy and PACE Overview}
\label{sec:overview}
Aggressive TCNOT scheduling reduces quantum execution time, but substantially increases the decoding workload. 
The conventional approach bounds this workload by separating consecutive TCNOTs by $d$ SE rounds, sacrificing much of the quantum-side speedup enabled by TCNOTs~\cite{sahay2025error}.
Our strategy is instead to reduce the decoding workload and schedule TCNOTs as densely as the available decoder resources permit.

Based on this strategy, we propose PACE, a framework that combines three decoder-side techniques with a decoder-aware TCNOT scheduler.
Sec.~\ref{sec:mitigation} presents the three complementary techniques for mitigating the bottlenecks, thereby increasing the TCNOT density that the decoder can sustain.
Sec.~\ref{sec:scheduling} presents a decoder-aware scheduler that constructs a high-density TCNOT schedule while satisfying decoder-side constraints.
In this way, PACE exploits the available decoder resources to increase TCNOT concurrency while preserving the FTQC pipeline.

\section{Mitigating Decoder-Side Bottlenecks}
\label{sec:mitigation}

\subsection{Hybrid Window Decoding}
\label{sec:hybrid}
As discussed in Sec.~\ref{sec:hyperedge}, TCNOTs can make window-level DEMs non-graphlike by introducing hyperedges.
A conservative approach is to use unmatchable decoders~\cite{beni2025tesseract,hillmann2025localized,muller2025improved} to process every window in a program containing TCNOTs. 
However, whether unmatchable decoders are actually required depends on the structure of each window-level DEM, rather than on the presence of TCNOTs in the program.

\begin{figure}[t]
    \centering
    \includegraphics[width=\linewidth]{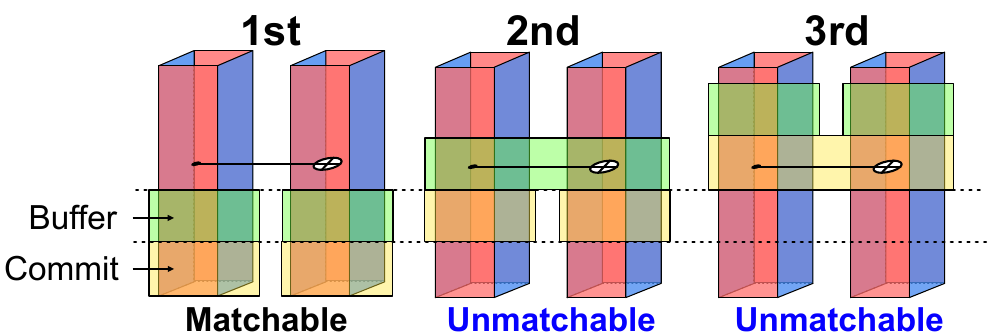}
    \caption{Per-window decoder selection in Hybrid window.}
    \label{fig:hybrid_window}
    
\end{figure}

\begin{keyobservation}
    \textbf{Key observation:}
    Many decoding windows in FTQC workloads with TCNOTs remain graphlike and can therefore be handled efficiently by matchable decoders.
\end{keyobservation}
We exploit this per-window heterogeneity through \emph{Hybrid window decoding}, which consists of two steps: \emph{per-window decoder selection} and \emph{window-wise correction}. 
Per-window decoder selection assigns a matchable decoder to graphlike DEMs and an unmatchable decoder otherwise, as shown in Fig.~\ref{fig:hybrid_window}.
Window-wise correction converts each decoder-specific fault estimate into a common logical Pauli-frame update and residual detection-event state for the next window.

For $i$-th window, given its DEM $D_i=(p_i,H_i,A_i)$ with $N_i$ fault mechanisms and detection-event vector $\sigma_i$, the decoder estimates a fault vector $\hat{e}_i\in\{0,1\}^{N_i}$ satisfying $H_i\hat e_i=\sigma_i$.
Matchable and unmatchable decoders estimate different sets of fault mechanisms; thus, $N_i$ and the dimensions of $H_i$ and $A_i$ may differ across windows.
Conventional window decoding assumes a common fault representation: it uses $\hat{e}_i$ to update $\sigma_{i+1}$ and eventually combines $\hat{e}_i$ to estimate a global observable. 
This procedure is not directly applicable when estimates produced by different decoder types do not share a common fault mechanism.
Window-wise correction instead consumes each estimate within the window, applying the corresponding action matrix before the estimate leaves its native fault space.

Let $\mathcal{D}_{C,i}$ be the set of detectors in the commit region of the $i$-th window.
We define the commit-region mask $m_{i} \in \{0,1\}^{N_i}$ as 
\begin{equation}
    (m_{i})_n = \bigvee_{k\in\mathcal{D}_{C,i}} (H_i)_{k,n},
    \label{eq:commit_mask}
\end{equation}
which selects the fault mechanisms that affect the commit region.
The corresponding logical correction and detection-event update are
\begin{align}
    \ell_{i} &= A_i\,(m_{i} \odot \hat{e}_i), \label{eq:logical_commit}\\
    \sigma_{i+1} &\leftarrow \sigma_i \oplus H_i\left(m_{i} \odot \hat{e}_i\right), \label{eq:detector_outcome_update}
\end{align}
where $\odot$ denotes element-wise product.
The logical corrections $\ell_{i}$ are accumulated in the logical Pauli frame, while the detection-event update for the next window removes the committed estimate. 
Thus, each decoder output is consumed in its native fault representation, preventing the decoder choice for one window from affecting subsequent windows.
This generalizes standard sliding-window decoding~\cite{gong2024toward} and extends naturally to parallel window decoding~\cite{skoric2023parallel}. 

\begin{figure*}[!t]
    \centering
    \includegraphics[width=0.7\linewidth]{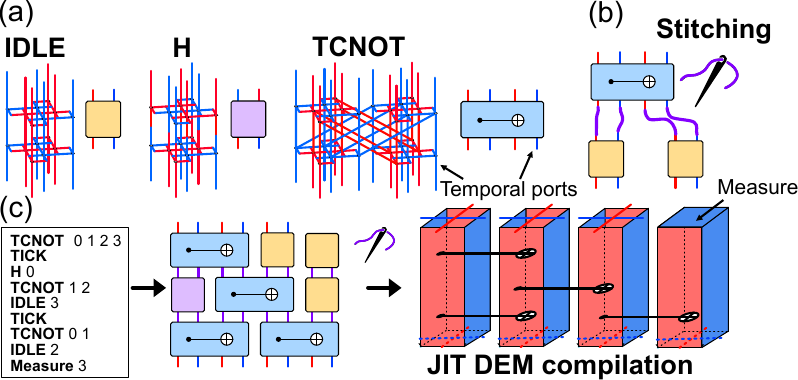}
    \caption{Overview of DEM Stitch:
    (a) an example of a local template,
    (b) stitching local templates along the time dimension,
    (c) template-based JIT construction of the window-level DEM.}
    \label{fig:dem_stitch}
\end{figure*}

\subsection{Just-in-Time (JIT) DEM Compilation}
\label{sec:dem_stitching}

As discussed in Sec.~\ref{sec:dem_patterns}, the window-level DEM depends on the TCNOT schedule, leading to exponentially many possible DEM patterns.
However, for online decoding, the DEM for each decoding window must be ready before its decoding task begins.
Because exhaustively precompiling all possible DEMs is infeasible, each schedule-specific DEM must be compiled just in time.

Existing methods construct each window-level DEM from its corresponding full circuit~\cite{gidney2021stim,ziad2026greenpeas}.
Applying this approach at runtime treats every schedule-specific DEM as an independent compilation target, even though different DEMs repeatedly contain similar local structures.

\begin{keyobservation}
    \textbf{Key observation:}
    Although TCNOT scheduling yields many window-level DEM patterns, each DEM is composed of a small set of recurring local circuit blocks whose local fault-to-detector structures can be reused across windows.
\end{keyobservation}

Based on this observation, we introduce \emph{DEM Stitch}, a template-based JIT DEM construction technique, which separates DEM compilation into two phases: \emph{offline template generation} and \emph{online template stitching}.
The offline phase precompiles recurring local blocks as reusable DEM templates, while the online phase instantiates and connects the templates according to the circuit and TCNOT schedule.

During offline template generation, DEM Stitch constructs reusable templates for local circuit blocks: initialization, measurement, idle, and logical Clifford operations, including TCNOT between two surface-code patches, as shown in Fig.~\ref{fig:dem_stitch}\,(a).
Each template captures spatial error propagation within one or two patches, while detector dependencies across temporal boundaries are exposed as \emph{temporal ports}.
These ports provide an interface through which neighboring template instances are connected during online assembly.
To track both $X$ and $Z$ logical observables within a single window, we generate the templates with a noiseless Bell-pair ancilla following~\cite{gidney2025yoked}.

During online template stitching, DEM Stitch scans the time blocks of each decoding window and instantiates the template corresponding to each circuit block.
For each template instance, it remaps local detector coordinates to global detector indices, propagates logical observables through subsequent TCNOT gates to determine their final indices, and appends the resulting fault columns to the window-level DEM.
It then resolves the temporal ports between adjacent template instances.
Algorithm~\ref{alg:dem_stitch} summarizes this procedure.

Thus, DEM Stitch replaces full-circuit DEM compilation with template instantiation, index remapping, logical-observable propagation, and temporal-port resolution.
By reusing precompiled local structures, DEM Stitch substantially reduces online JIT compilation overhead.
Moreover, any instruction sequence in Fig.~\ref{fig:logical_operations} can be supported using only a small set of reusable templates.

\begin{algorithm}[t]
\footnotesize
\caption{JIT DEM construction with DEM Stitch}
\label{alg:dem_stitch}
\begin{algorithmic}[1]
\Require TCNOT schedule $S$, current window $\mathcal{W}$, templates $\mathcal{T}$
\Ensure DEM $D=(p,H,A)$ for the current window
\State $\mathcal{F}\gets\emptyset$
\For{each time block $b\in\mathcal{W}$ in forward order}
    \State Identify patch groups $\mathcal{G}_b$ induced by active TCNOTs
    \Statex \hspace{2.2em}\textit{(e.g., TCNOT$(0,2)$: $\mathcal{G}_b=\bigl\{\{0,2\},\{1\}\bigr\}$)}
    \For{each group $g\in\mathcal{G}_b$}
        \State Select $T\in\mathcal{T}$ matching the local operations of $g$
        \State Remap detectors in $T$ to global detector rows
        \State Map logical observables through subsequent TCNOTs
        \State Append the remapped fault columns to $\mathcal{F}$
    \EndFor
\EndFor
\State Resolve temporal ports between adjacent template instances
\State Merge equivalent fault columns
\State \Return $D=(p,H,A)$ assembled from $\mathcal{F}$
\end{algorithmic}
\end{algorithm}

\begin{figure*}[t]
    \centering
    \includegraphics[width=0.6\linewidth]{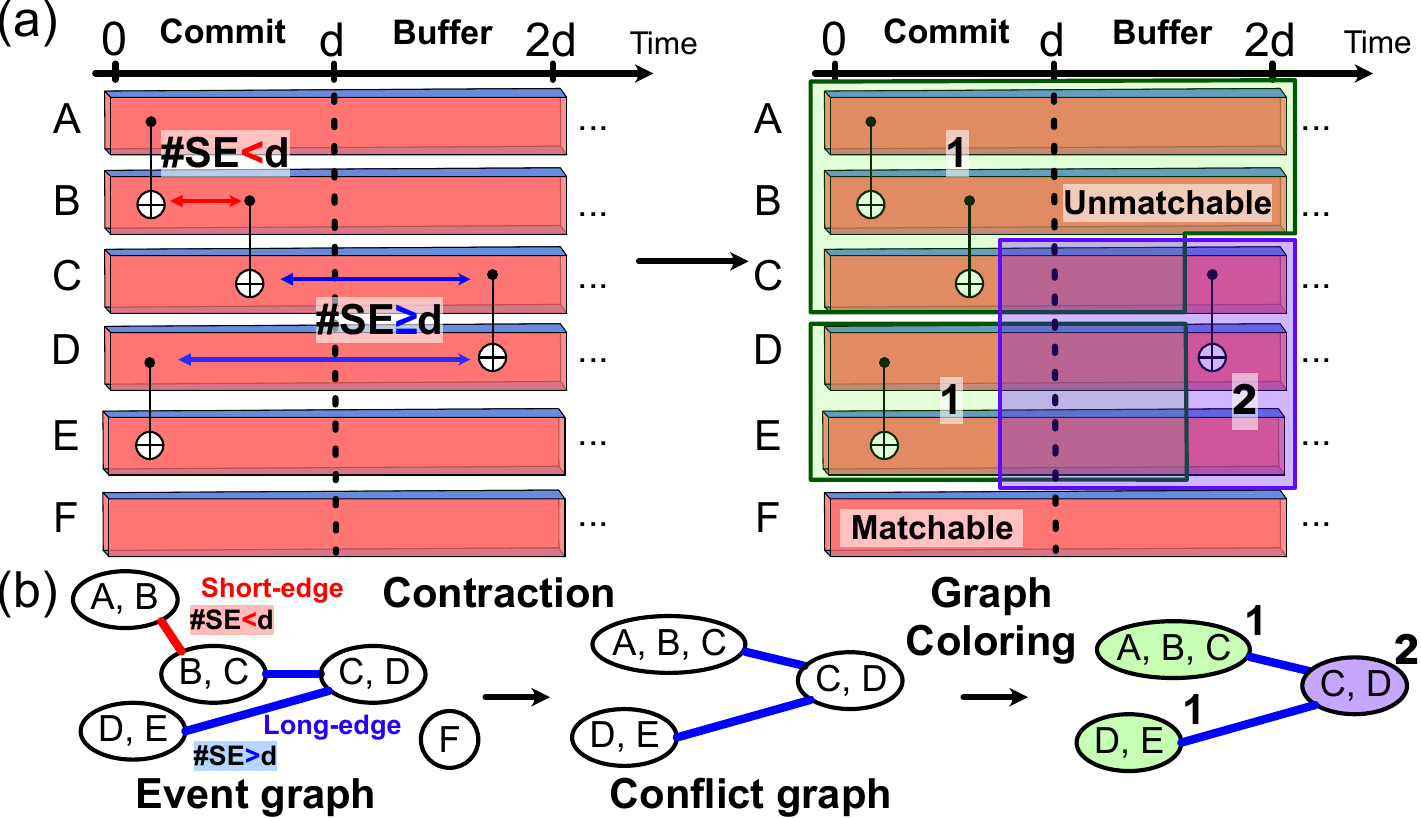}
    \caption{
        (a) Example of a height-$2d$ window with TCNOTs.
        SPD decomposes the window into smaller sub-windows when $\#\mathrm{SE}\ge d$ between consecutive TCNOT events on the same logical qubit.
        Sub-windows with the same color are decoded in parallel. 
        (b) Graph construction for SPD.
        Short-edge components in the event graph define the sub-windows, while long-edges between distinct sub-windows define the conflict graph.
        Coloring the conflict graph gives the parallel stages shown in (a).
    }
    \label{fig:graph_coloring}    
\end{figure*}

\subsection{Sub-window parallel decoding (SPD)}
\label{sec:graph_coloring}

As explained in Sec.~\ref{sec:window_decoding}, conventional sliding-window decoding processes each height-$W$ window as a single decoding task.
Under dense TCNOT scheduling, correlations across many surface-code patches can cause the resulting decoding task to exceed the memory capacity or real-time latency budget of a decoder instance, as discussed in Sec.~\ref{sec:decoding_volume}.

Dividing the window into smaller decoding tasks can reduce the per-task decoding volume, but the window cannot be divided arbitrarily due to the buffer requirement discussed in Sec.~\ref{sec:window_decoding}.
Importantly, the requirement constrains the amount of syndrome history retained \emph{after the end of each commit region}; it does not require every decoding window to have the same height $W$.

\begin{keyobservation}
    \textbf{Key observation:}
    A height-$W$ decoding window can be decomposed into variable-height, overlapping \emph{sub-windows}, provided that each commit region is followed by a buffer of $W_B=d$ rounds.
\end{keyobservation}

PACE exploits this flexibility through \emph{Sub-window parallel decoding (SPD)}, which applies the principle of parallel-window decoding~\cite{skoric2023parallel} within each conventional sliding window.
SPD forms smaller sub-windows whose commit regions collectively cover the original commit region and whose buffer regions may overlap.
For consecutive TCNOT events on the same logical qubit, a separation of at least $d$ SE rounds permits a sub-window boundary; otherwise, the events must be decoded jointly.

SPD consists of two steps: \emph{sub-window formation} and \emph{decoding-layer assignment}.
We formulate these steps as graph contraction and coloring, respectively.
The first step groups TCNOT events that must be decoded jointly into sub-windows, while the second assigns nonconflicting sub-windows to the same parallel decoding layer.
Once the sub-window boundaries are determined, DEM Stitch (Sec.~\ref{sec:dem_stitching}) efficiently assembles the DEM of each variable-height sub-window from reusable local templates.
The layers are committed in dependency order using the window-wise correction described in Sec.~\ref{sec:hybrid}.

We explain the two steps using the example in Fig.~\ref{fig:graph_coloring}.
Figure~\ref{fig:graph_coloring}(a) illustrates a height-$W$ decoding window over logical qubits $\mathbf{\mathsf{A}}$--$\mathbf{\mathsf{F}}$.
The horizontal axis denotes SE rounds, and only TCNOT events are shown for simplicity.

\noindent
\textbf{Sub-window formation: }
We first construct the \emph{event graph} shown on the left of Fig.~\ref{fig:graph_coloring}(b).
Each node represents a TCNOT event, and an edge connects each pair of consecutive TCNOT events involving the same logical qubit.
Let $\#\mathrm{SE}$ denote the number of SE rounds between the two endpoint events.
We classify an edge as a \emph{short-edge} if $\#\mathrm{SE}<d$ and as a \emph{long-edge} otherwise.
The endpoints of a short-edge cannot be separated while retaining the required $W_B=d$-round buffer and must therefore be decoded jointly.
Accordingly, each connected component formed by short-edges defines the TCNOT-correlated core of one sub-window, which is extended with the required commit and buffer regions.
In the example, this connected-component analysis decomposes the correlated region over $\mathbf{\mathsf{A}}$--$\mathbf{\mathsf{E}}$ into the three sub-windows shown on the right of Fig.~\ref{fig:graph_coloring}(a).
Note that qubit $\mathbf{\mathsf{F}}$ participates in no TCNOT and therefore remains an independent graphlike sub-window. 

\noindent
\textbf{Decoding-layer assignment: }
We next contract each connected component induced by short-edges into one node to obtain the \emph{conflict graph} shown on the right of Fig.~\ref{fig:graph_coloring}(b).
Each long-edge connecting two nodes induces a conflict edge between the corresponding sub-windows.
Although a long-edge permits a valid sub-window boundary, the two sub-windows have an ordered dependency along the same logical-qubit timeline and must therefore be assigned to different decoding layers.
A proper coloring of the conflict graph assigns conflicting sub-windows different colors, while sub-windows with the same color are decoded in parallel.
In the example, the conflict graph can be colored with two colors; hence, the three sub-windows are executed in the two decoding layers shown in Fig.~\ref{fig:graph_coloring}(a).
Because conflict-graph coloring is tightly coupled with TCNOT scheduling, its detailed procedure is presented as part of the TCNOT scheduling method in Sec.~\ref{sec:scheduling}.

\section{Decoder-Aware TCNOT Scheduling}
\label{sec:scheduling}
This section presents the decoder-aware TCNOT scheduling method of our PACE framework.
Its objective is to minimize quantum execution time while ensuring the FTQC pipeline in Fig.~\ref{fig:pipeline}.
We first establish a generic decoder feasibility model.
We then incorporate the three decoder-side techniques of Sec.~\ref{sec:mitigation} to derive the PACE-aware constraints evaluated for each decoding window.
Finally, we present the method to find TCNOT scheduling with a short quantum execution time that satisfies these constraints.

\subsection{Generic Decoder Feasibility Model}
This subsection defines a generic resource model for determining whether an individual decoding task can be processed online.
As discussed in Sec.~\ref{sec:decoding_volume}, denser TCNOT schedules can enlarge the correlated region of a decoding task, thereby increasing its DEM preparation latency, decoder working-memory requirement, and decoding latency.

To characterize the size of a decoding task, we define \emph{decoding volume} as $V=qw/W$, where $q$ is the number of logical patches covered by the task and $w$ is the window length. 
Thus, $V=1$ represents one surface-code patch over a height-$W$ window.
For fixed code distance $d$ and window height $W$, $V$ scales with the number of detectors in the task.

To translate the per-task cost into online feasibility conditions, we next consider the rate at which decoding tasks are generated.
Successive height-$W$ windows advance by the commit-region length $W_{\mathrm{C}}$.
Thus, we define their arrival interval as $\Delta=W_{\mathrm{C}}\tau_{\mathrm{SE}}$.
Let $M_{\mathrm{HW}}$ denote the working memory available to one decoder instance.

We model each performance component $r$ as a power-law function of $V$ as summarized in Table~\ref{tab:decoder_models}.
Assuming that DEM preparation and decoding are pipelined across successive windows, each stage must complete one task within $\Delta$\footnote{Strictly, backlog-free execution is determined by the syndrome generation interval and aggregate throughput of each stage, which depend on hardware parallelism and the number of available instances. To keep the model simple and hardware-agnostic, we instead impose the conservative sufficient condition that the per-task latency of each stage does not exceed $\Delta$.}, while each task’s working-memory must fit within $M_{\mathrm{HW}}$.

\begin{table}[t]
    \centering
    \caption{Performance components and their modeling}
    \label{tab:decoder_models}
    \begingroup
    \scriptsize
    \begin{tabular}{c|c|c}
        \toprule
        \textbf{Performance component} & \textbf{Constraint} & \textbf{Modeling with $V$}  \\
        \midrule
        DEM preparation latency $T_{\mathrm{DEM}}$
        &  $T_{\mathrm{DEM}} \leq \Delta$
        &$\alpha_{\mathrm{DEM}}V^{\gamma_{\mathrm{DEM}}}$
        \\
        Decoder working memory $M_{\mathrm{dec}}$
        & $M_{\mathrm{dec}} \leq M_\mathrm{HW}$
        & $\alpha_{\mathrm{mem}}V^{\gamma_{\mathrm{mem}}}$
        \\
        Decoding latency $T_{\mathrm{dec}}$
        & $T_{\mathrm{dec}} \leq \Delta$
        & $\alpha_{\mathrm{dec}}V^{\gamma_{\mathrm{dec}}}$
        \\
        \bottomrule
    \end{tabular}
    \endgroup
\end{table}

\subsection{PACE-Aware Feasibility Constraints}
\label{sec:constraint}

We next incorporate the three decoder-side techniques of Sec.~\ref{sec:mitigation} into the generic model above and derive the feasibility constraints used by the PACE scheduler.
Hybrid window decoding (Sec.~\ref{sec:hybrid}) assigns each task to a matchable or unmatchable decoder according to whether its DEM is graphlike.
Under PACE, every graphlike task assigned to a matchable decoder has a fixed decoding volume of $V=1$, and we assume these fixed-volume tasks do not become system bottlenecks. 
Thus, unless otherwise stated, the decoding volumes and task sets refer only to non-graphlike tasks processed by unmatchable decoders.
Moreover, DEM Stitch (Sec.~\ref{sec:dem_stitching}) accelerates DEM preparation; it reduces $\alpha_{\mathrm{DEM}}$.

SPD (Sec.~\ref{sec:graph_coloring}) changes both the volume of each decoding task and their parallel execution structure, which we model as follows.
For a given height-$W$ window with volume $V$, let $\mathcal S$ denote the set of non-graphlike sub-windows produced by SPD, and let $V_s \leq V$ denote the decoding volume of sub-window $s\in\mathcal S$.
Coloring the corresponding conflict graph produces the decoding layers $\mathcal{C} = \{\mathcal{C}_1, \mathcal{C}_2, \ldots,\mathcal{C}_K\}$, where sub-windows in the same layer are decoded in parallel.
Then, the latency of layer $\mathcal{C}_k$ is determined by its slowest sub-window, and the total decoding latency $T^{*}_{\mathrm{dec}}(\mathcal C)$ is the sum of these per-layer latencies.

Under the conservative feasibility model in Table~\ref{tab:decoder_models}, the constraints for each height-$W$ window become
\begin{subequations}
\label{eq:window_resource_constraints}
\begin{align}
    \max_{s\in\mathcal S}T_{\mathrm{DEM}}(V_s)
    &\le \Delta,
    \label{eq:window_prep}\\
    \max_{s\in\mathcal S}M_{\mathrm{dec}}(V_s)
    &\le M_{\mathrm{HW}},
    \label{eq:window_memory}\\
    T^{*}_{\mathrm{dec}}(\mathcal C)
    \equiv
    \sum_k\max_{s\in\mathcal C_k}T_{\mathrm{dec}}(V_s)
    &\le \Delta.
    \label{eq:window_decoding}
\end{align}
\end{subequations}
The first two constraints bound the largest per-sub-window DEM preparation latency and memory requirement.
The third bounds the total critical-path latency of the colored decoding layers.
These constraints assume sufficient resources to prepare all sub-window DEMs concurrently and to decode all sub-windows in the same layer in parallel.
Under this assumption, satisfying Eq.~\eqref{eq:window_resource_constraints} is sufficient to prevent the backlog problem.

\subsection{Decoder-Constrained TCNOT Scheduling}
We present the decoder-aware TCNOT scheduler of our PACE framework.
The spacing between consecutive TCNOTs controls the tradeoff between quantum execution time and decoder demand, as discussed in Sec.~\ref{sec:decoding_volume}.
Jointly optimizing all inter-TCNOT spacings is combinatorial because each spacing decision changes not only the quantum execution time, but also the SPD-induced sub-windows and conflict graphs.
Thus, PACE restricts the search to a bounded family of spacing policies, greedily constructs an earliest-start schedule for each policy, and selects the shortest candidate for which every height-$W$ decoding window satisfies Eq.~\eqref{eq:window_resource_constraints}.

For each $p$, PACE enforces a $d$-round separation at every $p$-th inter-TCNOT gap along each logical-qubit timeline and uses the minimum spacing permitted by circuit dependencies for the remaining gaps.
It greedily places each TCNOT at the earliest time satisfying the circuit dependencies and this spacing policy.
The enforced separations provide potential SPD boundaries; thus, increasing $p$ generally produces a denser and shorter schedule at the cost of greater decoder demand. 
The choice $p=1$ yields the All-$d$ baseline~\cite{sahay2025error}, while $p\rightarrow\infty$ yields the dependency-limited ASAP schedule.

For each candidate schedule $\mathbf{t}$, PACE constructs its height-$W$ decoding windows and applies SPD and Hybrid Window Decoding to obtain the set of non-graphlike sub-windows $\mathcal S_j$ for each window $j$.
It then obtains a coloring $\mathcal C_j$ using a decoding-latency-aware DSatur-based heuristic~\cite{brelaz1979new} that approximately minimizes $T_{\mathrm{dec}}^{*}(\mathcal C_j)$.
The candidate $\mathbf{t}$ is retained only if every decoding window satisfies Eq.~\eqref{eq:window_resource_constraints}.
Algorithm~\ref{alg:scheduling} summarizes this bounded search.

\begin{algorithm}[t]
\footnotesize
\caption{Decoder-constrained scheduling}
\label{alg:scheduling}
\begin{algorithmic}[1]

\Function{Feasible}{$p$}
\State $\mathbf t\leftarrow\Call{GenerateSchedule}{p}$
\For{each decoding window $j$ induced by $\mathbf t$}
\State $\mathcal S_j\leftarrow\Call{DecomposeSub-window}{j}$
\State ${\mathcal C_j}
\leftarrow\Call{Coloring}{\mathcal S_j}$
\Comment{Heuristically minimize $T_{\mathrm{dec}}^{*}$}
\If{window $j$ violates
Eq.~\eqref{eq:window_resource_constraints}}
\State \Return False
\EndIf
\EndFor
\State \Return True
\EndFunction

\If{$\Call{Feasible}{\text{inf.}}$}
\State \Return $\Call{GenerateSchedule}{\text{inf.}}$
\Comment{ASAP is already feasible}
\EndIf

\State $p_{\mathrm{lo}}\leftarrow 1$
\If{\textbf{not} $\Call{Feasible}{p_{\mathrm{lo}}}$}
\State \Return \textsc{Infeasible}
\Comment{All-$d$ is infeasible; no feasible schedule exists}
\EndIf

\State $p_{\mathrm{hi}}\leftarrow 2$
\While{$p_{\mathrm{hi}}\le p_{\max}$ and
$\Call{Feasible}{p_{\mathrm{hi}}}$}
\State $(p_{\mathrm{lo}},p_{\mathrm{hi}})
\leftarrow(p_{\mathrm{hi}},2p_{\mathrm{hi}})$
\EndWhile

\State $p^{}\leftarrow
\Call{FindLargestFeasible}
{p_{\mathrm{lo}},\min(p_{\mathrm{hi}},p_{\max}+1)}$
\State $\mathbf t^{}\leftarrow\Call{GenerateSchedule}{p^{}}$
\State \Return $\mathbf t^{}$

\end{algorithmic}
\end{algorithm}

\section{Evaluation Setup}
\label{sec:setup}

\subsection{Decoder Configuration}
We set $W=2d$ and $W_{\mathrm{C}}=d$ for all window-decoding-based experiments.
Following~\cite{zhou2025resourceanalysis}, we assume $\tau_{\mathrm{SE}}=1$ms, which yields $\Delta=d\tau_{\mathrm{SE}}$. 
Unless otherwise noted, we use a code distance of $d=23$ throughout the evaluation.
We assume sufficient resources to prepare all sub-window DEMs concurrently and enough decoder instances to process all sub-windows within each layer in parallel. 
All CPU-based measurements are performed on an AMD Ryzen Threadripper 9970X system with 128~GiB of memory.

\subsection{Normalized Decoder Capacities}
As shown in Eq.~\eqref{eq:window_resource_constraints}, sustaining real-time decoding requires satisfying three resource constraints.
To express all three constraints in a common form, we rewrite them as equivalent bounds on decoding volume.
We define $V_{\mathrm{mem}}$, $V_{\mathrm{DEM}}$, and $V_{\mathrm{dec}}$ by
$M_{\mathrm{dec}}(V_{\mathrm{mem}})=M_{\mathrm{HW}}$,
$T_{\mathrm{DEM}}(V_{\mathrm{DEM}})=\Delta$, and
$T_{\mathrm{dec}}(V_{\mathrm{dec}})=\Delta$, respectively.
Because DEM preparation and memory constrain each sub-window independently, we combine them into the per-sub-window capacity
$V_{\mathrm{unit}}=\min(V_{\mathrm{mem}},V_{\mathrm{DEM}})$.
Decoding latency, in contrast, accumulates across sequential decoding layers and is therefore bounded separately by $V_{\mathrm{dec}}$.
For positive scaling exponents, Eq.~\eqref{eq:window_resource_constraints} can be rewritten as
\begin{subequations}
\label{eq:normalized_window_constraint}
    \begin{align}
        \max_{s\in\mathcal S} V_{s}
        &\le V_{\mathrm{unit}},
        \label{eq:normalized_unit_constraint}\\
        T^{*}_{\mathrm{dec}}(\mathcal C) = 
            \sum_k
            \max_{s\in\mathcal C_{k}}
            \left (
            \alpha_{\mathrm{dec}}
            V_{s}^{\gamma_{\mathrm{dec}}}
            \right )
        & \le \alpha_{\mathrm{dec}} V_{\mathrm{dec}}^{\gamma_\mathrm{dec}}.
        \label{eq:normalized_decoding_constraint}
    \end{align}
\end{subequations}
for all windows.
Equation~\eqref{eq:normalized_unit_constraint} bounds the largest individual sub-window,
whereas Eq.~\eqref{eq:normalized_decoding_constraint} bounds the accumulated decoding latency with SPD.

\subsection{FTQC Benchmark workloads}
\label{sec:benchmarks}
We use 10 practical FTQC workloads spanning five representative algorithm classes with various logical-qubit scales, as summarized in Table~\ref{tab:benchmark_characterization}.
The four programs from FTCircuitBench~\cite{harkness2026ftcircuitbench} cover arithmetic, Quantum Fourier Transform (QFT), linear-algebra kernels, and quantum singular value transform (QSVT). 
For application workloads, we evaluate the SELECT subroutine of quantum phase estimation (QPE) for various Hamiltonians ~\cite{yoshioka2023hunting,boyd2023low}: Jellium, Fermi--Hubbard, and Heisenberg models.
By using the technique in \cite{yoshioka2023hunting}, we compile the same SELECT workload at two data-loading parallelism levels; the parenthesized number $n$ denotes $2^n$-way parallelism.
This comparison isolates how instruction-level parallelism changes TCNOT concurrency and decoder demand without changing the target application.
To characterize the decoder demand before applying PACE, $V_{\mathrm{ASAP}}$ reports the largest decoding volume under ASAP scheduling, with SPD disabled at $d=23$ and $W=2d$.
A larger $V_{\mathrm{ASAP}}$ therefore indicates a higher decoder demand under the ASAP schedule.

\begin{table}[t]
    \centering
    \caption{Benchmark circuit characteristics after Clifford+$T$ compilation. LQ: logical qubits; OP: operations.}
    \label{tab:benchmark_characterization}
    \scriptsize
    \renewcommand{\arraystretch}{0.8}
    \setlength{\tabcolsep}{3pt}
    \resizebox{\columnwidth}{!}{%
    \begin{tabular}{llrrrrrr}
        \toprule
        \textbf{Benchmark} & \textbf{Class} & \textbf{\#LQ} & \textbf{\#Ops} & \textbf{\#CNOT} & \textbf{\#S} & \textbf{\#T} & $\boldsymbol{V}_{\mathrm{ASAP}}$ \\
        \midrule
        add64 & Arith. & 64 & 1,023 & 455 & 0 & 392 & 70 \\
        qft29 & QFT & 29 & 296,196 & 812 & 60,966 & 116,775 & 656 \\
        hhl7  & Lin. sys. & 7 & 73,271 & 317 & 15,179 & 28,729 & 144 \\
        qsvt6 & QSVT & 6 & 403,527 & 2,132 & 89,861 & 154,586 & 77 \\
        \midrule
        Jel2(1)  & QPE & 36 & 23,888 & 7,536 & 2,912 & 2,560 & 48 \\
        Jel2(4)  & QPE & 316 & 25,076 & 8,084 & 2,984 & 2,848 & 600 \\
        FH8(1)   & QPE & 148 & 27,088 & 8,580 & 1,633 & 2,692 & 47 \\
        FH8(5)   & QPE & 700 & 30,068 & 9,844 & 1,816 & 3,424 & 1,040 \\
        Hei12(1) & QPE & 166 & 37,308 & 11,352 & 3,783 & 6,492 & 48 \\
        Hei12(6) & QPE & 1,356 & 43,210 & 14,010 & 4,136 & 7,904 & 2,616 \\
        \bottomrule
    \end{tabular}
    }
    
\end{table}

Rotation gates are synthesized into Clifford+$T$ circuits using GridSynth~\cite{ross2016optimal} with a synthesis tolerance of $10^{-10}$.
Logical $S$ and $T$ gates use the teleportation circuits in Fig.~\ref{fig:logical_operations}.
We assume worst-case conditional $S$ corrections and resource states $\ket{Y}$ and $\ket{T}$ are always available~\cite{litinski2019magic,gidney2024magic}. 

\section{Evaluation Results}
\label{sec:results}
We evaluate PACE in four stages.
First, we quantify the benefits of the three decoder-side mitigation techniques (Sec.~\ref{sec:results_hybrid}--\ref{sec:results_spatial}). 
Second, we evaluate the end-to-end benefit of the decoder-aware TCNOT scheduler (Sec.~\ref{sec:results_scheduling}). 
Third, we isolate the effects of the normalized decoder capacities and assess the sensitivity of the selected schedules to the decoding-latency scaling model (Sec.~\ref{sec:contract_sensitivity}). 
Finally, we characterize the bottlenecks of current decoder systems and estimate the improvements required for fast FTQC with TCNOTs (Sec.~\ref{sec:case_study}).

\begin{figure*}[!t]
    \centering
    \includegraphics[width=0.8\linewidth]{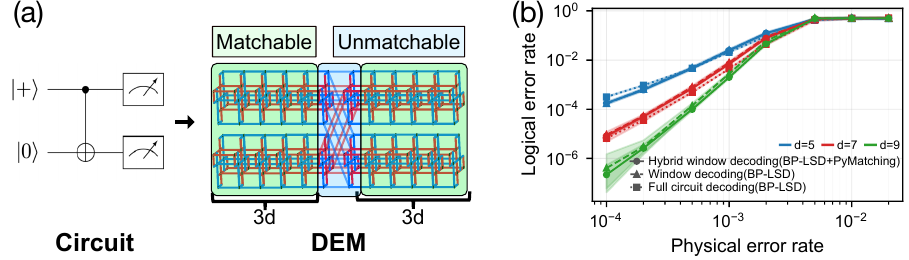}
    \caption{(a): Hybrid window decoding experiment using Bell-pair preparation with TCNOT.
    (b): Logical error rates for Hybrid window decoding.}
    \label{fig:ical_hybrid}
\end{figure*}

\subsection{Hybrid Window Decoding Preserves Accuracy}
\label{sec:results_hybrid}
We first verify that the Hybrid window decoding scheme introduced in Sec.~\ref{sec:hybrid} preserves the decoding accuracy.
We simulate the Bell-pair circuit with a TCNOT shown in Fig.~\ref{fig:ical_hybrid} using Stim~\cite{gidney2021stim} under a uniform circuit-level depolarizing noise model.
We use PyMatching~\cite{higgott2025sparse} and BP-LSD~\cite{hillmann2025localized} for the matchable and unmatchable decoders, respectively, and compare three configurations: (1)~Hybrid Window Decoding; 
(2)~window decoding with unmatchable decoder; and (3)~full-circuit decoding with unmatchable decoder. 
Comparing (1) and (2) isolates the effect of the per-window decoder selection, while comparing (2) and (3) validates the window-wise correction against full-circuit decoding.

As shown in Fig.~\ref{fig:ical_hybrid}, Hybrid Window Decoding achieves logical error rates consistent with both BP-LSD baselines.
These results show that Hybrid Window Decoding can route graphlike windows to the lower-cost matchable decoder without degrading logical error rate.

\begin{figure*}[t]
    \centering
    \includegraphics[width=0.8\linewidth]{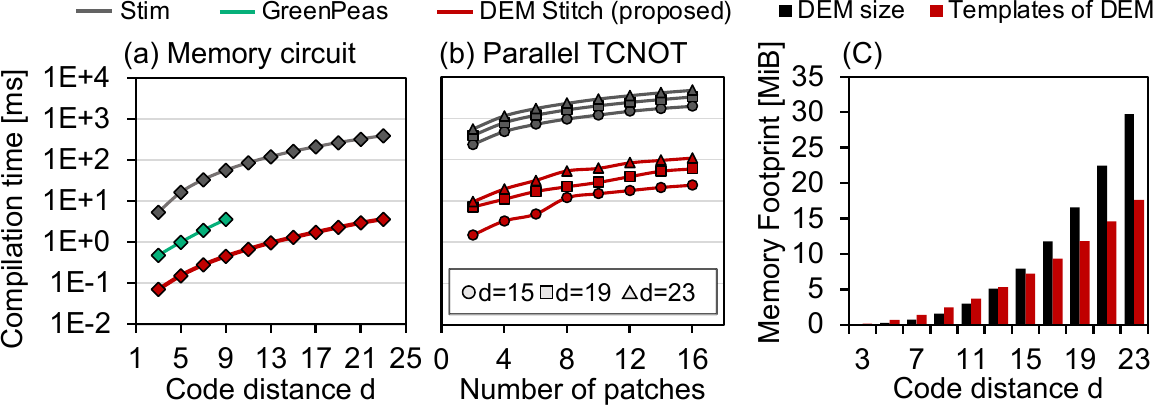}
    \caption{Online construction of schedule-specific DEMs. (a) Preparation time for a $W=2d$ memory window. (b) Preparation time versus the number of patches in a TCNOT window. (c) Template-cache footprint.}
    \label{fig:dem_stitch_jit_compile}    
\end{figure*}

\subsection{DEM Stitch Accelerates Online DEM Compilation}
\label{sec:results_stitch}
We next evaluate whether DEM Stitch can prepare schedule-specific DEMs within the online deadline and characterize its scaling with code distance and patch count.
We measure DEM-preparation latency for $W=2d$ memory experiments and for TCNOT windows spanning multiple surface-code patches.
The baseline invokes Stim's \texttt{detector\_error\_model()} on the complete window circuit, whereas DEM Stitch assembles the same DEM from cached templates.
We also report the template-cache footprint compared with that of full DEM with $W=2d$ memory experiment.
For context, we include the published GreenPeas results~\cite{ziad2026greenpeas}; because they were obtained using a different implementation and hardware platform, we treat them as a reference rather than a controlled baseline.

As shown in Fig.~\ref{fig:dem_stitch_jit_compile}(a), for a $d=23$ memory window, DEM Stitch is $108.4\times$ faster than full-circuit Stim construction.
Its measured latency is also lower than the reported GreenPeas values at the code distances evaluated by both studies.
At $d=23$, DEM Stitch meets the $\Delta=23$~ms deadline for the memory window and TCNOT windows spanning up to four patches, although larger correlated windows remain preparation-limited, as shown in Fig.~\ref{fig:dem_stitch_jit_compile}(a) and (b).
Over the evaluated range, its preparation latency scales approximately linearly with $d$.
The template-cache footprint remains modest over the evaluated code distances, reaching $17.5$~MiB at $d=23$, as shown in Fig.~\ref{fig:dem_stitch_jit_compile}(c).
The key implication is that template reuse removes most redundant construction work, but preparation must still be included in the decoding-system capacity contract.

\subsection{SPD Reduces Decoding Volume}
\label{sec:results_spatial}
We next isolate the volume-reduction effect of SPD from that of the scheduler.
We fix the spacing between consecutive TCNOTs on each logical qubit to $k\in\{1,2,\ldots,d\}$ SE rounds, and compare each schedule with and without SPD.

\begin{figure*}[p]
    \centering
    \includegraphics[width=0.8\linewidth]
        {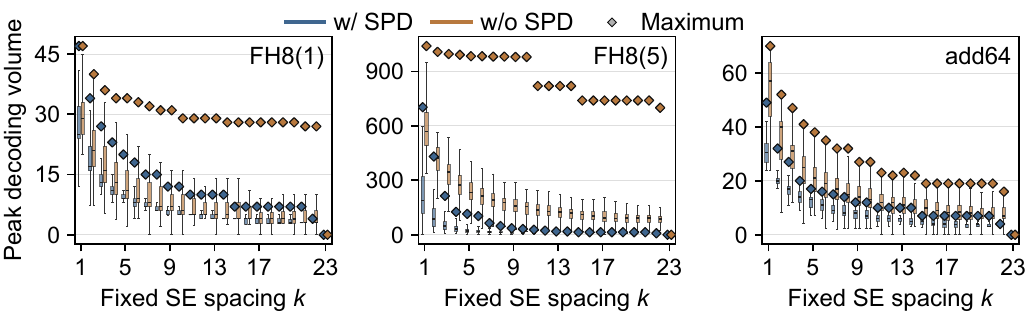}
    \caption{
        Peak decoding volume for fixed TCNOT spacing $k$, with and
        without SPD, at $d=23$.
        Each box summarizes the window distribution for the same schedule.
        The $k=1$ maximum without SPD equals $V_{\mathrm{ASAP}}$
        in Table~\ref{tab:benchmark_characterization}.
    }
    \vspace{5pt}
    \label{fig:fixed_k}
\end{figure*}

Figure~\ref{fig:fixed_k} reports the resulting distribution of peak decoding volume across active windows.
SPD shifts the per-window volume distributions downward for all three workloads, including a $3.0\times$ median reduction for the dense FH8(5) benchmark at $k=1$.
The benefit diminishes as $k$ approaches $d$ because temporal spacing itself increasingly provides valid window decomposition boundaries.
These results demonstrate that SPD is particularly effective at reducing decoding volume under dense TCNOT schedules.

\begin{figure*}[p]
    \centering
    \includegraphics[width=0.85\linewidth]
        {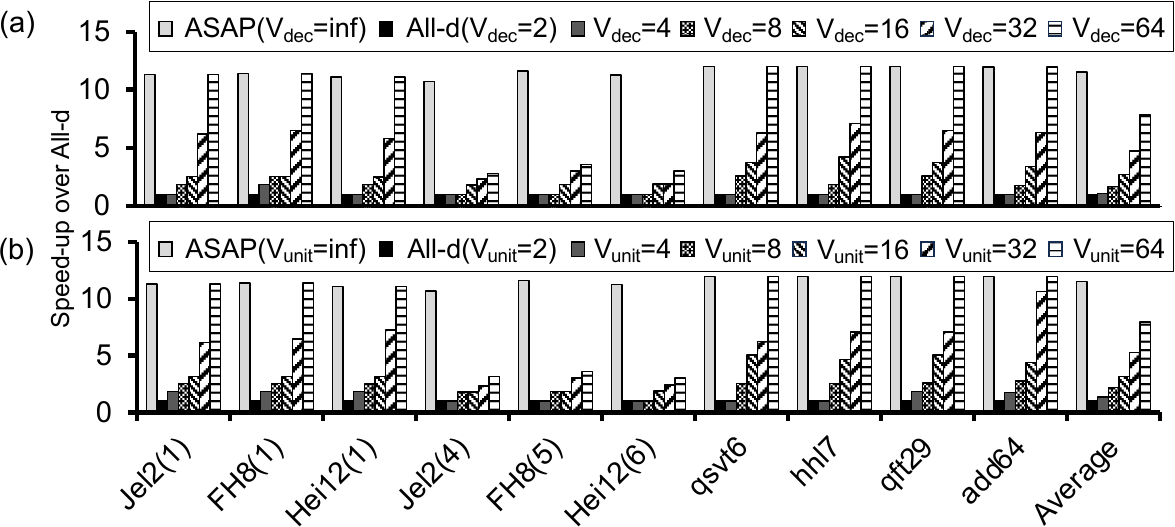}
    \caption{
        Quantum-execution speedup of schedules adaptively selected by
        PACE under independent capacity sweeps.
        Each result shows the shortest evaluated schedule satisfying both
        modeled decoder constraints.
        (a) $V_{\mathrm{dec}}$ is swept with $V_{\mathrm{unit}}$
        nonlimiting; (b) $V_{\mathrm{unit}}$ is swept with
        $V_{\mathrm{dec}}$ nonlimiting.
    }
    \vspace{5pt}
    \label{fig:scheduler_summary}
\end{figure*}

\begin{figure*}[p]
    \centering
    \includegraphics[width=0.8\linewidth]
        {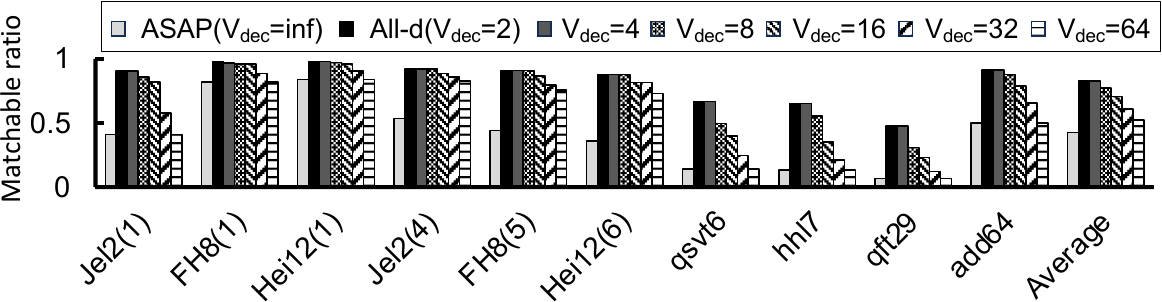}
    \caption{
        Per-workload ratio of graphlike to total decoding volume for
        the schedules selected by the $V_{\mathrm{dec}}$ sweep in
        Fig.~\ref{fig:scheduler_summary}(a).
    }
    \vspace{5pt}
    \label{fig:scheduler_matchable_volume_ratio}
\end{figure*}

\subsection{PACE Exploits Available Decoder Capacity}
\label{sec:results_scheduling}
Having evaluated the three decoder-side techniques individually, we next evaluate whether PACE can translate additional decoder capacity into shorter quantum schedules while satisfying the modeled constraints.
With SPD enabled, $d=23$, and $\gamma_{\mathrm{dec}}=1.25$, we independently sweep $V_{\mathrm{dec}}$ and $V_{\mathrm{unit}}$ over $\{4,8,16,32,64\}$ while setting the other capacity to a nonlimiting value.
This isolates the critical-path decoding constraint in Eq.~\eqref{eq:normalized_decoding_constraint} from the per-sub-window constraint in Eq.~\eqref{eq:normalized_unit_constraint}.
For each capacity setting, PACE selects the shortest evaluated schedule satisfying both constraints in Eq.~\eqref{eq:normalized_window_constraint}.
We report 
\clearpage
quantum-execution speedup over the All-$d$ baseline, including inserted SE rounds, with ASAP providing the capacity-unconstrained reference.

Figure~\ref{fig:scheduler_summary} shows that the selected schedules become progressively denser as decoder capacity increases.
At small capacities, only candidates with relatively wide TCNOT spacing satisfy the active constraint, so their execution times remain close to that of All-$d$.
Larger values of either $V_{\mathrm{dec}}$ or $V_{\mathrm{unit}}$ permit denser schedules and greater speedups while preserving decoder feasibility.
In the $V_{\mathrm{dec}}$ sweep, the geometric-mean speedup over All-$d$ increases from $1.06\times$ at $V_{\mathrm{dec}}=4$ to $7.86\times$ at $V_{\mathrm{dec}}=64$.
The $V_{\mathrm{unit}}$ sweep exhibits the same overall trend.

The required capacity also depends on workload structure.
The low-parallelism SELECT variants place fewer mutually correlated TCNOTs in each decoding window and therefore admit dense schedules at smaller capacities.
The high-parallelism variants place more TCNOTs within the same window, producing larger correlated regions and requiring greater decoder capacity before similarly dense schedules become feasible~\cite{sahay2025error,zhou2025resourceanalysis}.
PACE thus adapts TCNOT spacing to both the workload and the available decoder capacity.

When $V_{\mathrm{unit}}\ge V_{\mathrm{dec}}$, Eq.~\eqref{eq:normalized_decoding_constraint} subsumes Eq.~\eqref{eq:normalized_unit_constraint}.
We therefore focus the remaining analyses on this decoding-limited regime, where the selected schedule is determined solely by $V_{\mathrm{dec}}$.

Within this regime, increasing $V_{\mathrm{dec}}$ allows PACE to select denser TCNOT schedules, reducing the fraction of graphlike decoding windows.
Accordingly, Fig.~\ref{fig:scheduler_matchable_volume_ratio} shows that the mean graphlike share across workloads decreases from $82.7\%$ at $V_{\mathrm{dec}}=4$ to $52.3\%$ at $V_{\mathrm{dec}}=64$.
Nevertheless, even at the largest evaluated capacity, graphlike windows still account for more than half of the total decoding volume on average.
Hybrid window decoding can therefore continue to offload a substantial fraction of the workload to the matchable decoder while PACE exploits the additional decoder capacity.

\subsection{Sensitivity to Decoding-Time Scaling}
\label{sec:contract_sensitivity}

\begin{figure*}[t]
    \centering
    \includegraphics[width=0.75\linewidth]{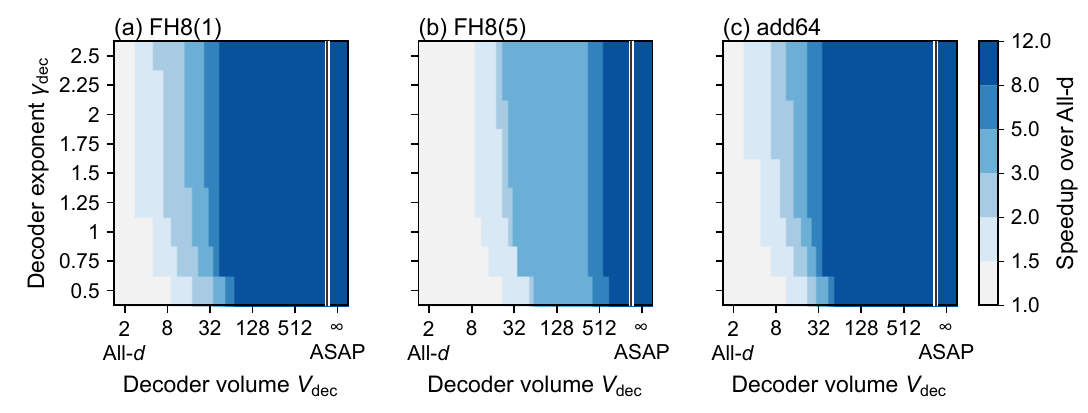}
    \caption{Effect of the decoding-time scaling exponent $\gamma_{\mathrm{dec}}$ on TCNOT scheduling across $V_{\mathrm{dec}}$ for three representative workloads.
    Each cell reports quantum-execution speedup over All-$d$ for the shortest evaluated schedule satisfying Eq.~\eqref{eq:normalized_decoding_constraint} at that or a smaller $V_{\mathrm{dec}}$.
    The $V_{\mathrm{dec}}=2$ column is the All-$d$ reference, and the separated $V_{\mathrm{dec}}=\infty$ column is the ASAP reference.}
    \label{fig:scheduler_contract_sensitivity}
\end{figure*}

We next examine how the decoding-latency exponent $\gamma_{\mathrm{dec}}$ affects schedule selection in the decoding-limited regime.
Figure~\ref{fig:scheduler_contract_sensitivity} sweeps $\gamma_{\mathrm{dec}}$ and $V_{\mathrm{dec}}$ for three representative workloads.
The $V_{\mathrm{dec}}=2$ and $V_{\mathrm{dec}}=\infty$ columns provide the All-$d$ and capacity-unconstrained ASAP references, respectively.

At a fixed $V_{\mathrm{dec}}$, changing $\gamma_{\mathrm{dec}}$ changes the relative benefit of SPD through Eq.~\eqref{eq:normalized_decoding_constraint}.
A larger $\gamma_{\mathrm{dec}}$ penalizes large decoding units more strongly, making decomposition increasingly beneficial despite the additional buffer overlap and sequential decoding layers.
Consequently, PACE generally selects denser TCNOT schedules as $\gamma_{\mathrm{dec}}$ increases.
This benefit saturates once the effective decoding volume is sufficiently small to satisfy Eq.~\eqref{eq:normalized_decoding_constraint}, because further reductions do not enable a shorter candidate schedule.
The transition points remain workload dependent, while the staircase boundaries arise from the discrete candidate schedules generated using the period $p$.

\begin{table}[t]
    \centering
    \small
    \caption{Representative configurations}
    \label{tab:case_study_setting}
    \begin{tabular}{@{}p{0.47\columnwidth}p{0.5\columnwidth}@{}}
        \toprule
        \textbf{Parameter} & \textbf{Configuration} \\
        \midrule
        Distance and window & $d\in\{15,19,23\}$; $W=2d$ \\
        Real-time deadline & $W_{\mathrm{C}}\tau_{\mathrm{SE}}=d\tau_{\mathrm{SE}}$; $\tau_{\mathrm{SE}}=1$~ms~\cite{zhou2025resourceanalysis} \\
        Unmatchable decoders & BP-LSD~\cite{hillmann2025localized}; Relay-BP~\cite{muller2025improved} \\
        Matchable decoders (reference) & PyMatching~\cite{higgott2025sparse} \\
        CPU memory & 4~GiB (assumed) \\
        FPGA iteration latency & 24~ns/complete iteration~\cite{maurer2025real} \\
        FPGA memory & 26.7~MiB (AMD VU19P FPGA) \\
        \bottomrule
    \end{tabular}    
     \vspace{-9pt}
\end{table}

\subsection{Case Study: Decoder Bottlenecks and Feasibility}
\label{sec:case_study}
Finally, we map representative decoder implementations to the normalized capacity model and identify the bottlenecks that limit fast TCNOT-based FTQC.
We measure decoding latency and resident working memory using one-patch surface-code memory windows with $W=2d$ and circuit-level noise $p=10^{-3}$. 
For each $d\in\{15,19,23\}$, we generate one-patch rotated surface code memory tasks with temporal extent $d\le w \le 3d$ rounds under circuit-level noise $p=10^{-3}$.
We use $W=2d$ as the normalization reference in the decoding volume $V=qw/W$.
For each setting, we average five timed runs after warm-up, excluding circuit construction, DEM generation, and syndrome sampling, and fit the power-law models in Table~\ref{tab:decoder_models}.
BP-LSD~\cite{hillmann2025localized} and Relay-BP~\cite{muller2025improved} represent unmatchable-decoder implementations, while PyMatching~\cite{higgott2025sparse} is included as a graphlike reference.
We extrapolate these measured volume-scaling models to the multi-patch decoding tasks induced by TCNOTs.
This volume-only approximation does not model additional dependence on hyperedge density or topology and should therefore be interpreted as a case study rather than a complete performance prediction.

Table~\ref{tab:case_study_setting} summarizes the evaluated configurations.
The CPU configurations use the measured fits and assume a $4$~GiB per-instance memory budget.
We derive $V_{\mathrm{DEM}}$ from the measured DEM Stitch performance in Sec.~\ref{sec:results_stitch}.
For the illustrative Relay-BP FPGA projection, we estimate latency from the reported $24$~ns per iteration~\cite{maurer2025real} and our measured iteration counts, assume linear scaling with decoding volume, and use a $26.7$~MiB memory budget, assuming a high-end FPGA.

\begin{figure*}[p]
    \centering
    \includegraphics[width=0.75\linewidth]{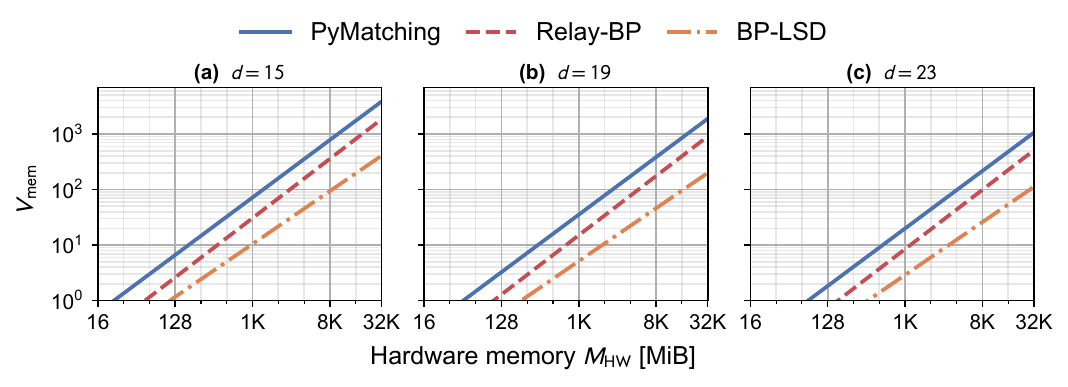}
    \caption{Memory-side capacity $V_{\mathrm{mem}}$ derived from the measured working-memory scaling of $W=2d$ decoding windows.}
    \label{fig:memory_sensitivity}
\end{figure*}

Figure~\ref{fig:memory_sensitivity} maps each memory budget to the corresponding capacity $V_{\mathrm{mem}}$.
PyMatching accommodates larger normalized volumes than either unmatchable decoder, while $V_{\mathrm{mem}}$ decreases with code distance for all three implementations.

\begin{figure*}[p]
    \centering
    \includegraphics[width=0.75\linewidth]{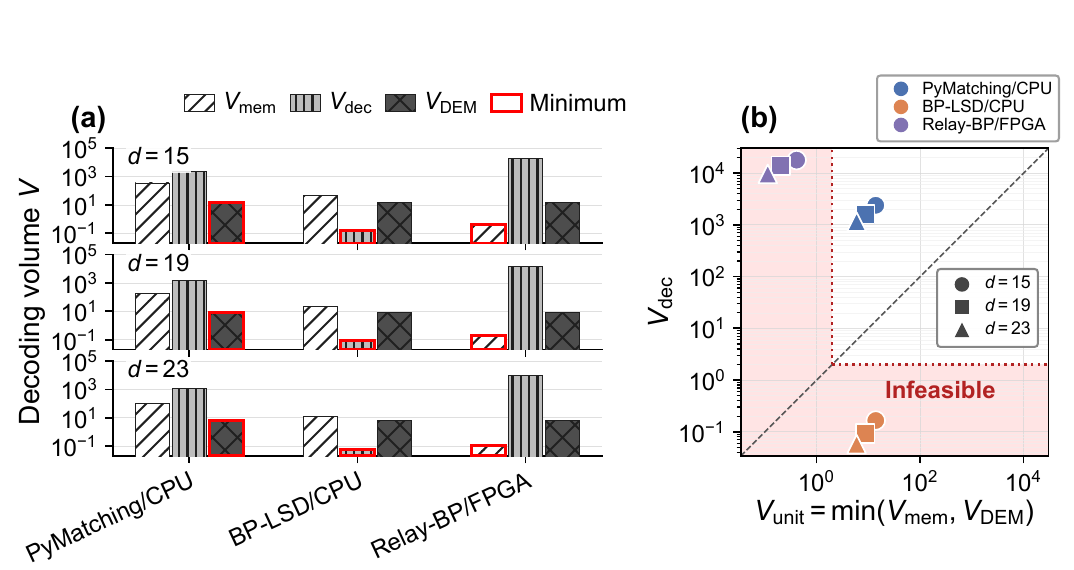}
    \caption{Capacities of the representative decoder configurations. 
    (a) DEM preparation, working-memory, and decoding latency capacities. 
    (b) The resulting $V_{\mathrm{unit}}$ and $V_{\mathrm{dec}}$.}
    \label{fig:case_study}
\end{figure*}

Figure~\ref{fig:case_study} compares the resulting preparation, memory, and decoding capacities.
CPU BP-LSD is limited by decoding latency, whereas the Relay-BP FPGA projection is limited by memory capacity.
None of the evaluated unmatchable-decoder configurations reaches the minimum capacities required even for the All-$d$ schedule, $V_{\mathrm{dec}}\ge2$ and $V_{\mathrm{unit}}\ge2$, corresponding to a two-patch TCNOT task.
In contrast, PyMatching satisfies both decoder-capacity constraints for graphlike decoding, leaving DEM preparation as the remaining bottleneck.

This case study demonstrates that the proposed decoding-volume model can identify both the real-time feasibility and the limiting resource of a decoder configuration.
For the evaluated systems, enabling fast FTQC based on TCNOT therefore requires faster unmatchable decoding, greater on-chip memory capacity for FPGA implementations, or faster JIT DEM preparation.
Alternatively, increasing $\tau_{\mathrm{SE}}$ relaxes the classical deadline, but proportionally increases quantum execution time.
\clearpage

\clearpage

\section{Discussion}
\label{sec:discussion}

\subsection{Finite Decoder Pools}
Our evaluation assumes enough unmatchable-decoder instances to process all non-graphlike sub-windows in each decoding layer concurrently, as well as sufficient matchable-decoder capacity for Hybrid Window Decoding.
With finite decoder pools, some nominally parallel tasks would be serialized, introducing decoder-assignment, contention, and queueing effects across decoding layers and successive windows.
Formally, the per-layer maximum in Eq.~\eqref{eq:window_decoding} would be replaced by the makespan of assigning the layer's sub-windows to the available decoder instances.

Extending PACE to this setting would therefore require jointly selecting the TCNOT schedule and assigning its decoding tasks to available decoders, similar to the decoder allocation problem studied by Triage~\cite{chen2026triage}.

\subsection{Complementarity with Fast Correlated Decoding}
PACE and the fast correlated decoding~\cite{cain2025fast} address complementary aspects of the decoding bottleneck caused by fast TCNOTs; PACE modifies the TCNOT schedule to control decoder workload, whereas fast correlated decoding decomposes each non-graphlike task induced by a given schedule into a sequence of graphlike tasks.
Although this approach accelerates decoding, it does not directly address the rapid growth in decoding volume and online DEM compilation of schedule-dependent DEMs, which our PACE framework addresses.
Fast correlated decoding could serve as the backend for non-graphlike sub-windows identified by SPD and Hybrid Window Decoding of our PACE framework.

Moreover, our case study in Sec.~\ref{sec:case_study} shows that PACE can accelerate TCNOT-based quantum execution even when all non-graphlike tasks are converted into graphlike tasks and decoded using PyMatching under the performance assumptions of current decoding systems. 
This result indicates that decoder-aware TCNOT scheduling remains beneficial even with a fast correlated-decoding backend.

\section{Related Work}
Window decoding has been widely studied~\cite{terhal2015quantum,toshio2025decoder,gong2024toward,lee2026efficient}, including the scheduling of dependent decoding tasks~\cite{chen2026triage,viszlai2025swiper,bombin2023modular,maurya2024managing}. Swiper~\cite{viszlai2025swiper} and Triage~\cite{chen2026triage} improve decoder utilization by scheduling window-level tasks, whereas PACE reduces the size and preparation cost of the correlated decoding tasks induced by dense TCNOT schedules.

\section{Conclusion}
\label{sec:conclusion}
In this work, we proposed the PACE framework, which leverages the three decoder-side techniques for mitigating the classical decoding bottleneck and enables decoder-aware scheduling for TCNOTs.
Our evaluations using practical FTQC benchmarks showed that PACE effectively utilizes available decoding resources to accelerate quantum program execution compared with baseline scheduling policies.
Furthermore, our case study with current decoding systems identifies the limitations that impede fast TCNOT-based FTQC and guides future decoder development.

\begin{acks}
This work was supported by MEXT Q-LEAP Grant No.~JPMXS0120319794, JPMXS0118068682, JST Moonshot R\&D Grant No.~JPMJMS256E, JPMJMS256L, JST CREST Grant No.~JPMJCR23I4, JPMJCR24I4, JPMJCR25I4, JSPS KAKENHI Grant No.~JP22H05000, JP25K21176.
\end{acks}

\bibliographystyle{ACM-Reference-Format}
\bibliography{references}

@article{terhal2015quantum,
  title = {Quantum error correction for quantum memories},
  author = {Terhal, Barbara M.},
  journal = {Reviews of Modern Physics},
  volume = {87},
  number = {2},
  pages = {307--346},
  year = {2015},
  doi = {10.1103/RevModPhys.87.307},
  publisher = {American Physical Society}
}

@article{dennis2002topological,
  title = {Topological quantum memory},
  author = {Dennis, Eric and Kitaev, Alexei and Landahl, Andrew and Preskill, John},
  journal = {Journal of Mathematical Physics},
  volume = {43},
  number = {9},
  pages = {4452--4505},
  year = {2002},
  doi = {10.1063/1.1499754},
  publisher = {AIP Publishing}
}

@article{horsman2012lattice,
  title = {Surface code quantum computing by lattice surgery},
  author = {Horsman, Clare and Fowler, Austin G. and Devitt, Simon and Van Meter, Rodney},
  journal = {New Journal of Physics},
  volume = {14},
  pages = {123011},
  year = {2012},
  doi = {10.1088/1367-2630/14/12/123011},
  publisher = {IOP Publishing}
}

@article{litinski2019game,
  title = {A game of surface codes: Large-scale quantum computing with lattice surgery},
  author = {Litinski, Daniel},
  journal = {Quantum},
  volume = {3},
  pages = {128},
  year = {2019},
  doi = {10.22331/q-2019-03-05-128},
  publisher = {Verein zur F{\"o}rderung des Open Access Publizierens in den Quantenwissenschaften}
}

@article{gidney2021stim,
  title = {Stim: a fast stabilizer circuit simulator},
  author = {Gidney, Craig},
  journal = {Quantum},
  volume = {5},
  pages = {497},
  year = {2021},
  doi = {10.22331/q-2021-07-06-497},
  publisher = {Verein zur F{\"o}rderung des Open Access Publizierens in den Quantenwissenschaften}
}

@misc{ziad2026greenpeas,
  title = {{GreenPeas}: Unlocking adaptive quantum error correction with just-in-time decoding hypergraphs},
  author = {Ziad, Abbas B. and Xu, Jubo and Fan, Hongxiang},
  year = {2026},
  eprint = {2604.16613},
  archivePrefix = {arXiv},
  primaryClass = {quant-ph},
  note = {arXiv:2604.16613},
}

@article{skoric2023parallel,
  title = {Parallel window decoding enables scalable fault tolerant quantum computation},
  author = {Skoric, Luka and Browne, Dan E. and Barnes, Kenton M. and Gillespie, Neil I. and Campbell, Earl T.},
  journal = {Nature Communications},
  volume = {14},
  number = {1},
  pages = {7040},
  year = {2023},
  doi = {10.1038/s41467-023-42482-1},
  publisher = {Springer Nature}
}

@article{higgott2025sparse,
  title = {Sparse Blossom: correcting a million errors per core second with minimum-weight matching},
  author = {Higgott, Oscar and Gidney, Craig},
  journal = {Quantum},
  volume = {9},
  pages = {1600},
  year = {2025},
  doi = {10.22331/q-2025-01-20-1600},
  publisher = {Verein zur F{\"o}rderung des Open Access Publizierens in den Quantenwissenschaften}
}

@article{hillmann2025localized,
  title = {Localized statistics decoding for quantum low-density parity-check codes},
  author = {Hillmann, Timo and Berent, Lucas and Quintavalle, Armanda O. and Eisert, Jens and Wille, Robert and Roffe, Joschka},
  journal = {Nature Communications},
  volume = {16},
  number = {1},
  pages = {8214},
  year = {2025},
  doi = {10.1038/s41467-025-63214-7},
  publisher = {Springer Nature}
}

@article{sahay2025error,
  title = {Error correction of transversal {CNOT} gates for scalable surface-code computation},
  author = {Sahay, Kaavya and Lin, Yingjia and Huang, Shilin and Brown, Kenneth R. and Puri, Shruti},
  journal = {PRX Quantum},
  volume = {6},
  number = {2},
  pages = {020326},
  year = {2025},
  doi = {10.1103/PRXQuantum.6.020326},
  publisher = {American Physical Society}
}

@article{cain2024correlated,
  title = {Correlated decoding of logical algorithms with transversal gates},
  author = {Cain, Madelyn and Zhao, Chen and Zhou, Hengyun and Meister, Nadine and Bonilla Ataides, J. Pablo and Jaffe, Arthur and Bluvstein, Dolev and Lukin, Mikhail D.},
  journal = {Physical Review Letters},
  volume = {133},
  number = {24},
  pages = {240602},
  year = {2024},
  doi = {10.1103/PhysRevLett.133.240602},
  publisher = {American Physical Society}
}

@misc{wan2024iterative,
  title = {An iterative transversal {CNOT} decoder},
  author = {Wan, Kwok Ho and Webber, Mark and Fowler, Austin G. and Hensinger, Winfried K.},
  year = {2024},
  eprint = {2407.20976},
  archivePrefix = {arXiv},
  primaryClass = {quant-ph},
  note = {arXiv:2407.20976},
}

@misc{sunami2025transversalgame,
  title = {Transversal surface-code game powered by neutral atoms},
  author = {Sunami, Shinichi and Goban, Akihisa and Yamasaki, Hayata},
  year = {2025},
  eprint = {2506.18979},
  archivePrefix = {arXiv},
  primaryClass = {quant-ph},
  note = {arXiv:2506.18979},
}

@article{zhou2025low,
  title = {Low-overhead transversal fault tolerance for universal quantum computation},
  author = {Zhou, Hengyun and Zhao, Chen and Cain, Madelyn and Bluvstein, Dolev and Maskara, Nishad and Duckering, Casey and Hu, Hongye and Wang, Sheng-Tao and Kubica, Aleksander and Lukin, Mikhail D.},
  journal = {Nature},
  volume = {646},
  number = {8084},
  pages = {303--308},
  year = {2025},
  doi = {10.1038/s41586-025-09543-5},
  publisher = {Springer Nature}
}

@inproceedings{zhou2025resourceanalysis,
  title = {Resource Analysis of Low-Overhead Transversal Architectures for Reconfigurable Atom Arrays},
  author = {Zhou, Hengyun and Duckering, Casey and Zhao, Chen and Bluvstein, Dolev and Cain, Madelyn and Kubica, Aleksander and Wang, Sheng-Tao and Lukin, Mikhail D.},
  booktitle = {Proceedings of the 52nd Annual International Symposium on Computer Architecture},
  pages = {1432--1448},
  year = {2025},
  publisher = {Association for Computing Machinery},
  doi = {10.1145/3695053.3731039},
}

@misc{chen2026triage,
  title = {Triage: An Adaptive Parallel Window Decoding Scheduler for Real-time Fault-Tolerant Quantum Computation},
  author = {Chen, Jiahan and Zhu, Chenghong and Bai, Ge and Wang, Xin},
  year = {2026},
  eprint = {2605.04459},
  archivePrefix = {arXiv},
  primaryClass = {quant-ph},
  note = {arXiv:2605.04459},
}

@misc{viszlai2024predictive,
  title = {Predictive Window Decoding for Fault-Tolerant Quantum Programs},
  author = {Viszlai, Joshua and Chadwick, Jason D. and Joshi, Sarang and Ravi, Gokul Subramanian and Li, Yanjing and Chong, Frederic T.},
  year = {2024},
  eprint = {2412.05115},
  archivePrefix = {arXiv},
  primaryClass = {quant-ph},
  note = {arXiv:2412.05115},
}

@article{chamberland2018pauli,
  title = {Fault-tolerant quantum computing in the {Pauli} or {Clifford} frame with slow error diagnostics},
  author = {Chamberland, Christopher and Iyer, Pavithran and Poulin, David},
  journal = {Quantum},
  volume = {2},
  pages = {43},
  year = {2018},
  doi = {10.22331/q-2018-01-04-43},
  publisher = {Verein zur F{\"o}rderung des Open Access Publizierens in den Quantenwissenschaften}
}

@article{bombin2023logical,
  title = {Logical blocks for fault-tolerant topological quantum computation},
  author = {Bombin, Hector and Dawson, Chris and Mishmash, Ryan V. and Nickerson, Naomi and Pastawski, Fernando and Roberts, Sam},
  journal = {PRX Quantum},
  volume = {4},
  pages = {020303},
  year = {2023},
  doi = {10.1103/PRXQuantum.4.020303},
  publisher = {American Physical Society}
}

@misc{cain2025fast,
  title = {Fast correlated decoding of transversal logical algorithms},
  author = {Cain, Madelyn and Bluvstein, Dolev and Zhao, Chen and Gu, Shouzhen and Maskara, Nishad and Kalinowski, Marcin and Geim, Alexandra A. and Kubica, Aleksander and Lukin, Mikhail D. and Zhou, Hengyun},
  year = {2025},
  eprint = {2505.13587},
  archivePrefix = {arXiv},
  primaryClass = {quant-ph},
  note = {arXiv:2505.13587},
}

@misc{harkness2026ftcircuitbench,
  title = {{FTCircuitBench}: A Benchmark Suite for Fault-Tolerant Quantum Compilation and Architecture},
  author = {Harkness, Adrian and Kan, Shuwen and Liu, Chenxu and Wang, Meng and Martyn, John M. and Xu, Shifan and Chamaki, Diana and Decker, Ethan and Mao, Ying and Zuluaga, Luis F. and Terlaky, Tam{\'a}s and Li, Ang and Stein, Samuel},
  year = {2026},
  eprint = {2601.03185},
  archivePrefix = {arXiv},
  primaryClass = {quant-ph},
  note = {arXiv:2601.03185},
}

@article{yoshioka2023hunting,
  title = {Hunting for quantum-classical crossover in condensed matter problems},
  author = {Yoshioka, Nobuyuki and Okubo, Tsuyoshi and Suzuki, Yasunari and Koizumi, Yuki and Mizukami, Wataru},
  journal = {npj Quantum Information},
  volume = {10},
  number = {1},
  pages = {45},
  year = {2024},
  doi = {10.1038/s41534-024-00839-4},
  publisher = {Springer Nature}
}

@article{fowler2012surface,
  title={Surface codes: Towards practical large-scale quantum computation},
  volume={86},
  ISSN={1094-1622},
  number={3},
  journal={Physical Review A},
  publisher={American Physical Society (APS)},
  author={Fowler, Austin G. and Mariantoni, Matteo and Martinis, John M. and Cleland, Andrew N.},
  year={2012},
  pages={032324},
  doi={10.1103/PhysRevA.86.032324}
}

@misc{gottesman2010introduction,
  title={An introduction to quantum error correction and fault-tolerant quantum computation},
  ISBN={9780821892848},
  ISSN={0160-7634},
  journal={Proceedings of Symposia in Applied Mathematics},
  publisher={American Mathematical Society},
  author={Gottesman, Daniel},
  year={2010},
  pages={13--58},
  doi={10.1090/psapm/068/2762145}
}

@misc{bravyi1998quantum,
  author={S. B. Bravyi and A. Yu. Kitaev},
  title={Quantum codes on a lattice with boundary},
  year={1998},
  eprint={quant-ph/9811052},
  archivePrefix={arXiv},
  primaryClass={quant-ph},
  note={arXiv:quant-ph/9811052},
}

@article{bluvstein2025fault,
  title={A fault-tolerant neutral-atom architecture for universal quantum computation},
  volume={649},
  ISSN={1476-4687},
  number={8095},
  journal={Nature},
  publisher={Springer Science and Business Media LLC},
  author={Bluvstein, Dolev and Geim, Alexandra A. and Li, Sophie H. and Evered, Simon J. and Bonilla Ataides, J. Pablo and Baranes, Gefen and Gu, Andi and Manovitz, Tom and Xu, Muqing and Kalinowski, Marcin and Majidy, Shayan and Kokail, Christian and Maskara, Nishad and Trapp, Elias C. and Stewart, Luke M. and Hollerith, Simon and Zhou, Hengyun and Gullans, Michael J. and Yelin, Susanne F. and Greiner, Markus and Vuletić, Vladan and Cain, Madelyn and Lukin, Mikhail D.},
  year={2025},
  pages={39--46},
  doi={10.1038/s41586-025-09848-5}
}

@article{bluvstein2023logical,
  title={Logical quantum processor based on reconfigurable atom arrays},
  volume={626},
  ISSN={1476-4687},
  number={7997},
  journal={Nature},
  publisher={Springer Science and Business Media LLC},
  author={Bluvstein, Dolev and Evered, Simon J. and Geim, Alexandra A. and Li, Sophie H. and Zhou, Hengyun and Manovitz, Tom and Ebadi, Sepehr and Cain, Madelyn and Kalinowski, Marcin and Hangleiter, Dominik and Bonilla Ataides, J. Pablo and Maskara, Nishad and Cong, Iris and Gao, Xun and Sales Rodriguez, Pedro and Karolyshyn, Thomas and Semeghini, Giulia and Gullans, Michael J. and Greiner, Markus and Vuletić, Vladan and Lukin, Mikhail D.},
  year={2023},
  pages={58--65},
  doi={10.1038/s41586-023-06927-3}
}

@misc{beni2025tesseract,
  author={Laleh Aghababaie Beni and Oscar Higgott and Noah Shutty},
  title={Tesseract: A Search-Based Decoder for Quantum Error Correction},
  year={2025},
  eprint={2503.10988},
  archivePrefix={arXiv},
  primaryClass={quant-ph},
  note={arXiv:2503.10988},
}

@article{roffe2020decoding,
  author={Joschka Roffe and David R. White and Simon Burton and Earl T. Campbell},
  title={Decoding Across the Quantum LDPC Code Landscape},
  journal={Physical Review Research},
  volume={2},
  number={4},
  pages={043423},
  year={2020},
  doi={10.1103/PhysRevResearch.2.043423},
  eprint={2005.07016},
  archivePrefix={arXiv},
  primaryClass={quant-ph}
}

@misc{muller2025improved,
  author={Tristan Müller and Thomas Alexander and Michael E. Beverland and Markus Bühler and Blake R. Johnson and Thilo Maurer and Drew Vandeth},
  title={Improved belief propagation is sufficient for real-time decoding of quantum memory},
  year={2025},
  eprint={2506.01779},
  archivePrefix={arXiv},
  primaryClass={quant-ph},
  note={arXiv:2506.01779},
}

@misc{beverland2025fail,
  author={Michael E. Beverland and Malcolm Carroll and Andrew W. Cross and Theodore J. Yoder},
  title={Fail fast: techniques to probe rare events in quantum error correction},
  year={2025},
  eprint={2511.15177},
  archivePrefix={arXiv},
  primaryClass={quant-ph},
  note={arXiv:2511.15177},
}

@misc{fowler2018low,
  author={Austin G. Fowler and Craig Gidney},
  title={Low overhead quantum computation using lattice surgery},
  year={2018},
  eprint={1808.06709},
  archivePrefix={arXiv},
  primaryClass={quant-ph},
  note={arXiv:1808.06709},
}

@article{delfosse2021almost,
  title={Almost-linear time decoding algorithm for topological codes},
  volume={5},
  ISSN={2521-327X},
  journal={Quantum},
  publisher={Verein zur Forderung des Open Access Publizierens in den Quantenwissenschaften},
  author={Delfosse, Nicolas and Nickerson, Naomi H.},
  year={2021},
  pages={595},
  doi={10.22331/q-2021-12-02-595}
}

@article{griffiths2024union,
  title={Union-find quantum decoding without union-find},
  volume={6},
  ISSN={2643-1564},
  number={1},
  journal={Physical Review Research},
  publisher={American Physical Society (APS)},
  author={Griffiths, Sam J. and Browne, Dan E.},
  year={2024},
  pages={013154},
  doi={10.1103/PhysRevResearch.6.013154}
}

@article{cicali2025fast,
  title={Fast neutral-atom transport and transfer between optical tweezers},
  volume={24},
  ISSN={2331-7019},
  number={2},
  journal={Physical Review Applied},
  publisher={American Physical Society (APS)},
  author={Cicali, Cristina and Calzavara, Martino and Cuestas, Eloisa and Calarco, Tommaso and Zeier, Robert and Motzoi, Felix},
  year={2025},
  pages={024070},
  doi={10.1103/7r3w-8m61}
}

@inproceedings{riesebos2017pauli,
  series={DAC ’17},
  title={Pauli Frames for Quantum Computer Architectures},
  booktitle={Proceedings of the 54th Annual Design Automation Conference 2017},
  publisher={ACM},
  author={Riesebos, L. and Fu, X. and Varsamopoulos, S. and Almudever, C. G. and Bertels, K.},
  year={2017},
  pages={1--6},
  doi={10.1145/3061639.3062300},
  collection={DAC ’17}
}

@article{litinski2019magic,
  title={Magic State Distillation: Not as Costly as You Think},
  volume={3},
  ISSN={2521-327X},
  journal={Quantum},
  publisher={Verein zur Forderung des Open Access Publizierens in den Quantenwissenschaften},
  author={Litinski, Daniel},
  year={2019},
  pages={205},
  doi={10.22331/q-2019-12-02-205}
}

@misc{gidney2024magic,
  author={Craig Gidney and Noah Shutty and Cody Jones},
  title={Magic state cultivation: growing T states as cheap as CNOT gates},
  year={2024},
  eprint={2409.17595},
  archivePrefix={arXiv},
  primaryClass={quant-ph},
  note={arXiv:2409.17595},
}

@article{gidney2025yoked,
  title={Yoked surface codes},
  volume={16},
  ISSN={2041-1723},
  number={1},
  journal={Nature Communications},
  publisher={Springer Science and Business Media LLC},
  author={Gidney, Craig and Newman, Michael and Brooks, Peter and Jones, Cody},
  year={2025},
  pages={4498},
  doi={10.1038/s41467-025-59714-1}
}

@misc{gong2024toward,
  author={Anqi Gong and Sebastian Cammerer and Joseph M. Renes},
  title={Toward Low-latency Iterative Decoding of QLDPC Codes Under Circuit-Level Noise},
  year={2024},
  eprint={2403.18901},
  archivePrefix={arXiv},
  primaryClass={quant-ph},
  note={arXiv:2403.18901},
}

@inproceedings{viszlai2025swiper,
  title={Swiper: Minimizing fault-tolerant quantum program latency via speculative window decoding},
  author={Viszlai, Joshua and Chadwick, Jason D and Joshi, Sarang and Ravi, Gokul Subramanian and Li, Yanjing and Chong, Frederic T},
  booktitle={Proceedings of the 52nd Annual International Symposium on Computer Architecture},
  pages={1386--1401},
  year={2025},
  publisher={ACM},
  doi={10.1145/3695053.3731022}
}

@article{ross2016optimal,
  title={Optimal ancilla-free Clifford+T approximation of z-rotations},
  volume={16},
  ISSN={1533-7146},
  number={11\&12},
  journal={Quantum Information and Computation},
  publisher={Rinton Press},
  author={Ross, Neil J. and Selinger, Peter},
  year={2016},
  pages={901--953},
  doi={10.26421/QIC16.11-12-1}
}

@book{Nielsen_Chuang_2010,
  place={Cambridge},
  title={Quantum Computation and Quantum Information: 10th Anniversary Edition},
  publisher={Cambridge University Press},
  author={Nielsen, Michael A. and Chuang, Isaac L.},
  year={2010},
  doi={10.1017/CBO9780511976667}
}

@article{acharya2025belowthreshold,
  title = {Quantum error correction below the surface code threshold},
  author = {Acharya, Rajeev and Abanin, Dmitry A. and Aghababaie-Beni, Laleh and Aleiner, Igor and Andersen, Trond I. and Ansmann, Markus and Arute, Frank and Arya, Kunal and Asfaw, Abraham and Astrakhantsev, Nikita and Atalaya, Juan and Babbush, Ryan and Bacon, Dave and Ballard, Brian and Bardin, Joseph C. and Bausch, Johannes and Bengtsson, Andreas and Bilmes, Alexander and Blackwell, Sam and Boixo, Sergio and Bortoli, Gina and Bourassa, Alexandre and Bovaird, Jenna and Brill, Leon and Broughton, Michael and Browne, David A. and Buchea, Brett and Buckley, Bob B. and Buell, David A. and Burger, Tim and Burkett, Brian and Bushnell, Nicholas and Cabrera, Anthony and Campero, Juan and Chang, Hung-Shen and Chen, Yu and Chen, Zijun and Chiaro, Ben and Chik, Desmond and Chou, Charina and Claes, Jahan and Cleland, Agnetta Y. and Cogan, Josh and Collins, Roberto and Conner, Paul and Courtney, William and Crook, Alexander L. and Curtin, Ben and Das, Sayan and Davies, Alex and De Lorenzo, Laura and Debroy, Dripto M. and Demura, Sean and Devoret, Michel and Di Paolo, Agustin and Donohoe, Paul and Drozdov, Ilya and Dunsworth, Andrew and Earle, Clint and Edlich, Thomas and Eickbusch, Alec and Elbag, Aviv Moshe and Elzouka, Mahmoud and Erickson, Catherine and Faoro, Lara and Farhi, Edward and Ferreira, Vinicius S. and Burgos, Leslie Flores and Forati, Ebrahim and Fowler, Austin G. and Foxen, Brooks and Ganjam, Suhas and Garcia, Gonzalo and Gasca, Robert and Genois, {\'E}lie and Giang, William and Gidney, Craig and Gilboa, Dar and Gosula, Raja and Dau, Alejandro Grajales and Graumann, Dietrich and Greene, Alex and Gross, Jonathan A. and Habegger, Steve and Hall, John and Hamilton, Michael C. and Hansen, Monica and Harrigan, Matthew P. and Harrington, Sean D. and Heras, Francisco J. H. and Heslin, Stephen and Heu, Paula and Higgott, Oscar and Hill, Gordon and Hilton, Jeremy and Holland, George and Hong, Sabrina and Huang, Hsin-Yuan and Huff, Ashley and Huggins, William J. and Ioffe, Lev B. and Isakov, Sergei V. and Iveland, Justin and Jeffrey, Evan and Jiang, Zhang and Jones, Cody and Jordan, Stephen and Joshi, Chaitali and Juhas, Pavol and Kafri, Dvir and Kang, Hui and Karamlou, Amir H. and Kechedzhi, Kostyantyn and Kelly, Julian and Khaire, Trupti and Khattar, Tanuj and Khezri, Mostafa and Kim, Seon and Klimov, Paul V. and Klots, Andrey R. and Kobrin, Bryce and Kohli, Pushmeet and Korotkov, Alexander N. and Kostritsa, Fedor and Kothari, Robin and Kozlovskii, Borislav and Kreikebaum, John Mark and Kurilovich, Vladislav D. and Lacroix, Nathan and Landhuis, David and Lange-Dei, Tiano and Langley, Brandon W. and Laptev, Pavel and Lau, Kim-Ming and Le Guevel, Lo{\"i}ck and Ledford, Justin and Lee, Joonho and Lee, Kenny and Lensky, Yuri D. and Leon, Shannon and Lester, Brian J. and Li, Wing Yan and Li, Yin and Lill, Alexander T. and Liu, Wayne and Livingston, William P. and Locharla, Aditya and Lucero, Erik and Lundahl, Daniel and Lunt, Aaron and Madhuk, Sid and Malone, Fionn D. and Maloney, Ashley and Mandr{\`a}, Salvatore and Manyika, James and Martin, Leigh S. and Martin, Orion and Martin, Steven and Maxfield, Cameron and McClean, Jarrod R. and McEwen, Matt and Meeks, Seneca and Megrant, Anthony and Mi, Xiao and Miao, Kevin C. and Mieszala, Amanda and Molavi, Reza and Molina, Sebastian and Montazeri, Shirin and Morvan, Alexis and Movassagh, Ramis and Mruczkiewicz, Wojciech and Naaman, Ofer and Neeley, Matthew and Neill, Charles and Nersisyan, Ani and Neven, Hartmut and Newman, Michael and Ng, Jiun How and Nguyen, Anthony and Nguyen, Murray and Ni, Chia-Hung and Niu, Murphy Yuezhen and O'Brien, Thomas E. and Oliver, William D. and Opremcak, Alex and Ottosson, Kristoffer and Petukhov, Andre and Pizzuto, Alex and Platt, John and Potter, Rebecca and Pritchard, Orion and Pryadko, Leonid P. and Quintana, Chris and Ramachandran, Ganesh and Reagor, Matthew J. and Redding, John and Rhodes, David M. and Roberts, Gabrielle and Rosenberg, Eliott and Rosenfeld, Emma and Roushan, Pedram and Rubin, Nicholas C. and Saei, Negar and Sank, Daniel and Sankaragomathi, Kannan and Satzinger, Kevin J. and Schurkus, Henry F. and Schuster, Christopher and Senior, Andrew W. and Shearn, Michael J. and Shorter, Aaron and Shutty, Noah and Shvarts, Vladimir and Singh, Shraddha and Sivak, Volodymyr and Skruzny, Jindra and Small, Spencer and Smelyanskiy, Vadim and Smith, W. Clarke and Somma, Rolando D. and Springer, Sofia and Sterling, George and Strain, Doug and Suchard, Jordan and Szasz, Aaron and Sztein, Alex and Thor, Douglas and Torres, Alfredo and Torunbalci, M. Mert and Vaishnav, Abeer and Vargas, Justin and Vdovichev, Sergey and Vidal, Guifre and Villalonga, Benjamin and Heidweiller, Catherine Vollgraff and Waltman, Steven and Wang, Shannon X. and Ware, Brayden and Weber, Kate and Weidel, Travis and White, Theodore and Wong, Kristi and Woo, Bryan W. K. and Xing, Cheng and Yao, Z. Jamie and Yeh, Ping and Ying, Bicheng and Yoo, Juhwan and Yosri, Noureldin and Young, Grayson and Zalcman, Adam and Zhang, Yaxing and Zhu, Ningfeng and Zobrist, Nicholas},
  journal = {Nature},
  volume = {638},
  number = {8052},
  pages = {920--926},
  year = {2025},
  doi = {10.1038/s41586-024-08449-y},
  issn = {1476-4687}
}

@article{chen2026transversal,
  title={Transversal logical Clifford gates on the rotated surface code with reconfigurable neutral atom arrays},
  author={Chen, Zi-Han and Chen, Ming-Cheng and Lu, Chao-Yang and Pan, Jian-Wei},
  journal={Physical Review Letters},
  volume={136},
  number={13},
  pages={130601},
  year={2026},
  publisher={APS},
  doi={10.1103/m7tq-9v3g}
}

@misc{chen2026hierarchical,
  author={Zi-Han Chen and Ming-Cheng Chen and Chao-Yang Lu and Jian-Wei Pan},
  title={Hierarchical Logical Processor on the Rotated Surface Code with Shuttle Buses},
  year={2026},
  eprint={2606.22594},
  archivePrefix={arXiv},
  primaryClass={quant-ph},
  note={arXiv:2606.22594},
}

@article{brelaz1979new,
  title={New methods to color the vertices of a graph},
  volume={22},
  ISSN={1557-7317},
  number={4},
  journal={Communications of the ACM},
  publisher={Association for Computing Machinery (ACM)},
  author={Brélaz, Daniel},
  year={1979},
  pages={251--256},
  doi={10.1145/359094.359101}
}

@misc{toshio2025decoder,
  author={Riki Toshio and Kaito Kishi and Jun Fujisaki and Hirotaka Oshima and Shintaro Sato and Keisuke Fujii},
  title={Decoder Switching: Breaking the Speed-Accuracy Tradeoff in Real-Time Quantum Error Correction},
  year={2025},
  eprint={2510.25222},
  archivePrefix={arXiv},
  primaryClass={quant-ph},
  note={arXiv:2510.25222},
}

@article{battistel2023real,
  title={Real-time decoding for fault-tolerant quantum computing: progress, challenges and outlook},
  volume={7},
  ISSN={2399-1984},
  number={3},
  journal={Nano Futures},
  publisher={IOP Publishing},
  author={Battistel, F and Chamberland, C and Johar, K and Overwater, R W J and Sebastiano, F and Skoric, L and Ueno, Y and Usman, M},
  year={2023},
  pages={032003},
  doi={10.1088/2399-1984/aceba6}
}

@article{gidney2022stability,
  title={Stability Experiments: The Overlooked Dual of Memory Experiments},
  volume={6},
  ISSN={2521-327X},
  journal={Quantum},
  publisher={Verein zur Forderung des Open Access Publizierens in den Quantenwissenschaften},
  author={Gidney, Craig},
  year={2022},
  pages={786},
  doi={10.22331/q-2022-08-24-786}
}

@misc{maurer2025real,
  author={Thilo Maurer and Markus Bühler and Michael Kröner and Frank Haverkamp and Tristan Müller and Drew Vandeth and Blake R. Johnson},
  title={Real-time decoding of the gross code memory with FPGAs},
  year={2025},
  eprint={2510.21600},
  archivePrefix={arXiv},
  primaryClass={quant-ph},
  note={arXiv:2510.21600},
}

@misc{boyd2023low,
  title={Low-overhead parallelisation of lcu via commuting operators},
  author={Boyd, Gregory},
  archivePrefix={arXiv},
  eprint={2312.00696},
  year={2023},
  primaryClass={quant-ph},
  note={arXiv:2312.00696},
}

@inproceedings{wang2026transpiler,
  title={Transpiler-Architecture Co-Design to Curb Clifford Costs in Fault-Tolerant Quantum Computing},
  author={Wang, Meng and Liu, Chenxu and Stein, Samuel and Ding, Yufei and Das, Poulami and Nair, Prashant J and Li, Ang},
  booktitle={2026 ACM/IEEE 53rd Annual International Symposium on Computer Architecture (ISCA)},
  pages={889--904},
  year={2026},
  doi={10.1109/ISCA66397.2026.00072},
  organization={IEEE}
}

@article{lee2026efficient,
  title={Efficient post-selection for general quantum LDPC Codes},
  volume={12},
  ISSN={2056-6387},
  number={1},
  pages = {96},
  journal={npj Quantum Information},
  publisher={Springer Science and Business Media LLC},
  author={Lee, Seok-Hyung and English, Lucas H. and Bartlett, Stephen D.},
  doi = {10.1038/s41534-026-01242-x},
  year={2026}
}

@misc{bombin2023modular,
  author={Héctor Bombín and Chris Dawson and Ye-Hua Liu and Naomi Nickerson and Fernando Pastawski and Sam Roberts},
  title={Modular decoding: parallelizable real-time decoding for quantum computers},
  year={2023},
  eprint={2303.04846},
  archivePrefix={arXiv},
  primaryClass={quant-ph},
  note = {arXiv:2303.04846},
}

@misc{maurya2024managing,
  author={Satvik Maurya and Abtin Molavi and Aws Albarghouthi and Swamit Tannu},
  title={Managing Classical Processing Requirements for Quantum Error Correction},
  year={2024},
  eprint={2406.17995},
  archivePrefix={arXiv},
  primaryClass={quant-ph},
  note = {arXiv:2406.17995},
}

\end{document}